\documentclass[aps,prd,a4paper,reprint,nofootinbib,superscriptaddress,floatfix,preprintnumbers]{revtex4-2}

\usepackage{graphicx}
\usepackage{dcolumn}
\usepackage{bm}
\usepackage{hyperref}
\usepackage{todonotes}
\usepackage[version=4]{mhchem}
\usepackage{subcaption}
\usepackage[T1]{fontenc}
\usepackage{textcomp}
\usepackage{adjustbox}
\usepackage{multirow}
\usepackage[normalem]{ulem}

\newcommand{\lcdm}{$\Lambda \text{CDM}$}
\newcommand{\Neff}{\ensuremath{N_\mathrm{eff}}}

\newcommand{\GeV}{\mbox{GeV}}
\newcommand{\MeV}{\mbox{MeV}}
\newcommand{\keV}{\mbox{keV}}

\begin{document}


\title{Early-universe constraints on the electron mass}

\author{Michela Garramone}
\email{michela.garramone@unito.it}
\affiliation{%
University of Turin, Physics Department, via P.\ Giuria, 1, 10125 Turin (Italy)
}%
\affiliation{%
Istituto Nazionale di Fisica Nucleare (INFN), Turin section, via P.\ Giuria, 1, 10125 Turin (Italy)
}%
\affiliation{%
Instituto de Fisica Corpuscular (IFIC), CSIC-UV, c.\ Catedr\'atico J.\ Beltr\'an 2, 46980 Paterna - Valencia (Spain)
}%

\author{Nicolao Fornengo}
\email{nicolao.fornengo@unito.it}
\affiliation{%
University of Turin, Physics Department, via P.\ Giuria, 1, 10125 Turin (Italy)
}%
\affiliation{%
Istituto Nazionale di Fisica Nucleare (INFN), Turin section, via P.\ Giuria, 1, 10125 Turin (Italy)
}%

\author{Stefano Gariazzo}
\email{stefano.gariazzo@unito.it}
\affiliation{%
University of Turin, Physics Department, via P.\ Giuria, 1, 10125 Turin (Italy)
}%
\affiliation{%
Istituto Nazionale di Fisica Nucleare (INFN), Turin section, via P.\ Giuria, 1, 10125 Turin (Italy)
}%
\affiliation{%
Instituto de Fisica Corpuscular (IFIC), CSIC-UV, c.\ Catedr\'atico J.\ Beltr\'an 2, 46980 Paterna - Valencia (Spain)
}%

\date{\today}

\begin{abstract}
We investigate the impact of a nonstandard electron mass $m_e$ on early-Universe thermal history, focusing on neutrino decoupling and Big Bang Nucleosynthesis (BBN).
In the standard cosmology, neutrino--electron interactions keep neutrinos in thermal contact with the electromagnetic plasma until shortly before $e^\pm$ annihilation. Varying $m_e$ shifts the decoupling epoch and the entropy transfer from $e^\pm$ annihilation, thereby modifying the neutrino energy density and the inferred effective number of relativistic species, $N_{\mathrm{eff}}$.
Independently, during BBN the rates of charged-current weak processes, and hence the neutron-to-proton ratio, depend on $m_e$.
By confronting BBN predictions for the primordial light-element abundances with observations and imposing cosmological constraints on $N_{\mathrm{eff}}$, 
we obtain the following $1\sigma$ bounds on $m_e$ in the early Universe: $m_e = 0.505^{+0.006}_{-0.007}$ MeV (for the NACRE II nuclear reaction network) or $m_e=0.509^{+0.005}_{-0.004}$ MeV (for the PRIMAT nuclear reaction network). These bounds have been derived by adopting the recent determination of the primordial Helium-4 abundance by the Large Binocular Telescope observations of 54 metal-poor  H\,\textsc{ii} regions. If instead we adopt the Particle Data Book Helium-4 abundance, the bounds are: $m_e = 0.503^{+0.011}_{-0.015}$ MeV (NACRE II) or $m_e=0.521^{+0.009}_{-0.007}$ MeV (PRIMAT) The obtained allowed ranges are close to the present laboratory value at the level of $\sim 0.4\%-2\%$, depending on the dataset and nuclear network, thus supporting the constancy of the electron mass over cosmological timescales.
\end{abstract}

\maketitle

\section{\label{sec:intro}Introduction}

In the early universe, the epoch when the plasma had a temperature around the MeV scale is extremely interesting.
At that time, the universe is radiation-dominated and the particles that mostly constitute the thermal plasma are photons, neutrinos and electrons, with a small fraction of protons and neutrons~\cite{Lesgourgues:2018ncw}.
The interactions among these particles are central to this manuscript.

Within the \lcdm\ framework (see, e.g., Ref.~\cite{Peebles:2024txt}), one can compute key cosmological observables, including the Cosmic Microwave Background (CMB) anisotropy spectra, the matter power spectrum, and the primordial light-element abundances produced during Big Bang Nucleosynthesis (BBN)~\cite{Grohs:2023voo,Cooke:2024nqz}. These predictions have been extensively tested. In particular, the CMB has been measured with high precision by \emph{Planck}~\cite{Planck:2018nkj,Planck:2018vyg}, yielding some of the tightest constraints on the parameters of the standard cosmological model.

It is usually assumed that fundamental constants and symmetries are invariant throughout cosmic history, but it is well motivated to test this assumption observationally, see for example~\cite{Yoo:2002vw,Uzan:2024ded,Baryakhtar:2024rky, Baryakhtar:2025uxs,Barenboim:2026ocd}.
As an example, variations of the electron mass $m_e$ relative to its laboratory value have been studied in Refs.~\cite{Seto:2024cgo,Toda:2025dzd,Toda:2024ncp,Toda:2025kcq,Smith:2025icl}. Those works showed that $m_e$ affects Thomson scattering and hence the CMB anisotropy spectra, implying that $m_e$ at recombination must be close to its present value, with allowed deviations at the percent level (depending on the adopted datasets and cosmological model extensions).

In this work we probe earlier times, when the temperature of the electromagnetic plasma was comparable to $m_e$, i.e.\ at the MeV scale.
This epoch encompasses both neutrino decoupling and BBN. As we discuss below, each process is sensitive to $m_e$, providing complementary constraints on the electron mass in the early Universe.
We also stress that the electron mass effects on BBN have been partially discussed in Ref.~\cite{Seto:2022xgx}, although with a focus on the solution of the Hubble tension.
Let us emphasize that in our analysis we investigate the effect of a variation of $m_e$ only and we do not concurrently vary other fundamental constants.

\section{\label{sec:nudec}Neutrino decoupling}
In the standard hot Big Bang cosmology supplemented by the Standard Model, at photon temperatures $T\sim\mathcal{O}(10)\,\MeV$ the energy density is dominated by a nearly ultrarelativistic QED plasma of photons and electrons/positrons, together with three generations of neutrinos and antineutrinos. At these temperatures, frequent weak and electromagnetic interactions maintain all species close to thermal equilibrium and hence at a common temperature. As the Universe expands and cools to $T\sim\mathcal{O}(1)\,\MeV$, the interaction rate of neutrinos with the $e^\pm$ plasma falls below the Hubble expansion rate $H$, and neutrinos decouple.

After neutrino decoupling, the electromagnetic plasma continues to evolve through $e^+e^- \rightarrow \gamma\gamma$ annihilations. When the temperature drops to $T\sim m_e$, most $e^\pm$ pairs annihilate, transferring entropy to the photon bath. This entropy injection heats photons relative to the decoupled neutrinos, generating a temperature ratio $T_\gamma/T_\nu \simeq (11/4)^{1/3}\simeq 1.401$ in the instantaneous-decoupling limit. In the Standard Model the ratio receives small corrections because neutrino decoupling is not instantaneous.

Assuming equilibrium Bose--Einstein (photons) and Fermi--Dirac (neutrinos) distributions, the temperature ratio can be used to define the effective number of relativistic species, \Neff:
\begin{equation}
    \Neff \equiv 3\left(\frac{11}{4}\right)^{4/3}\left(\frac{T_\nu}{T_\gamma}\right)^4.
\end{equation}
In the Standard Model, accounting for noninstantaneous decoupling, its momentum dependence, and neutrino oscillations, one finds $\Neff\simeq 3.044$~\cite{Akita:2020szl,Froustey:2020mcq,Bennett:2020zkv}, consistent with current observational constraints~\cite{Planck:2018vyg}.

To compute \Neff\ in the presence of entropy production, one must solve a coupled system of evolution equations, including the continuity equation for the total energy density.
The energy density and pressure of species $i$ are
\begin{align}
    \rho_i &= \frac{g_i}{(2\pi)^3}\int E(\vec p)\,f(\vec p)\,d^3p , \label{density}\\
    P_i &= \frac{g_i}{(2\pi)^3}\int \frac{|\vec p|^2}{3E(\vec p)}\,f(\vec p)\,d^3p , \label{press}\\
    f(\vec p) &= \frac{1}{e^{E(\vec p)/T}\pm 1},
\end{align}
where $f(\vec p)$ is the Fermi--Dirac ($+$) or Bose--Einstein ($-$) distribution function (with vanishing chemical potential), $g_i$ counts internal degrees of freedom of particle species $i$, and $d^3p = 4\pi p^2\,dp$.

The continuity equation is usually written as
\begin{equation}
\frac{d\rho}{dt}=-3H(\rho+P),
\label{eq:continuity_phys}
\end{equation}
where $\rho$ and $P$ are the total energy density and pressure, and $H$ is the Hubble rate.
Neutrino-decoupling calculations are often performed in terms of comoving variables as functions of the scale factor $a(t)$,
\begin{equation}
x \equiv m_e a,\qquad y \equiv p a,\qquad z \equiv T a,
\label{eq:comoving}
\end{equation}
with $a(t)$ normalized such that $aT\to 1$ at sufficiently high temperatures. Defining comoving energy density and pressure as $\bar\rho\equiv a^4\rho$ and $\bar P\equiv a^4P$, Eq.~\eqref{eq:continuity_phys} becomes
\begin{equation}\label{eq:continuity}
x\,\frac{d\bar{\rho}}{dx}=\bar{\rho}-3\bar{P}.
\end{equation}

Here we instead adopt a fixed mass scale $m_0$ to define the comoving time variable,
\begin{equation}
x' \equiv m_0 a,\qquad x = k\,x',\qquad k \equiv \frac{m_e}{m_0},
\end{equation}
so that variations in $m_e$ do not change the definition of the independent variable $x'$.

An additional ingredient in the thermodynamics of the plasma is provided by finite-temperature QED corrections (FTQED), discussed, e.g., in Refs.~\cite{Fornengo:1997wa,Bennett:2019ewm}. These effects modify the energy density and pressure of the electromagnetic plasma,
\begin{align}
\bar{\rho} &= \sum_{i=\gamma,\nu_i,e,\mu} \bar{\rho}_i + \delta\bar{\rho},\\
\bar{P} &= \sum_{i=\gamma,\nu_i,e,\mu} \bar{P}_i + \delta\bar{P},
\end{align}
where $\delta\bar{\rho}$ and $\delta\bar{P}$ encode the QED interactions among photons and charged leptons at finite temperature (see, e.g., Ref.~\cite{Bennett:2019ewm} for explicit expressions).

Combining Eq.~\eqref{eq:continuity} with the expressions for $\rho_i$ and $P_i$ relevant at these temperatures (neutrinos, electrons/positrons, photons; see Appendix~\ref{sec:appendixA}), we obtain the evolution equation for the comoving photon temperature:
\begin{equation}\label{eq:dzodxp}
\frac{dz}{dx'}
=
\frac{
k^2\frac{x'}{z}J_2 - \frac{1}{2 z^3} \sum_{\alpha}\frac{d\bar\rho_{\nu_\alpha}}{dx'} + G_1(x',z)
}{k^2\frac{x'^2}{z^2}J_2 + J_4 + \frac{2}{15}\pi^2 + G_2(x',z)}\,,
\end{equation}
where $J_{2,4}$, $G_1(x',z)$, and $G_2(x',z)$ are defined in Appendix~\ref{sec:appendixA}, and $\bar\rho_{\nu_\alpha}$ is the comoving energy density of neutrinos of flavor $\alpha$.

\subsection{\label{sec:liouville}The Liouville equation}
After decoupling, neutrinos are not, in general, characterized by the same temperature as photons. Out of equilibrium, the evolution of the neutrino phase-space distribution is described by the kinetic (Boltzmann--Liouville) equation,
\begin{equation}\label{lio}
\frac{\partial f}{\partial t} - Hp\,\frac{\partial f}{\partial p} = \mathcal{C}[f],
\end{equation}
where $p$ is the physical momentum, $H$ is the Hubble rate, and $\mathcal{C}[f]$ is the collision term. In a homogeneous and isotropic Universe,
\begin{equation}
H^2 = \frac{8\pi G}{3}\,\rho_T = \frac{8\pi}{3}\frac{\rho_T}{m_{\rm pl}^2},
\end{equation}
where $\rho_T$ is the total energy density, $G$ is Newton's constant, and $m_{\rm pl}=1.22\times 10^{19}\,\GeV$ is the Planck mass.

Multiplying Eq.~\eqref{lio} by $gE$ and integrating over momentum space yields the energy-transfer form of the continuity equation for a species with $g$ internal degrees of freedom,
\begin{equation}
    \frac{d\rho}{dt} +3H(\rho + P) = \frac{\delta \rho}{\delta t}
    = \int \frac{g\,E\,d^3p}{(2\pi)^3}\,\mathcal{C}[f],
    \label{coneq}
\end{equation}
where $\delta\rho/\delta t$ is the net energy transfer rate between the species and the rest of the plasma. Summing Eq.~\eqref{coneq} over all species reproduces Eq.~\eqref{eq:continuity_phys}, since the collision terms cancel in the total.

Using $d\rho/dt = (dT/dt)\,(\partial\rho/\partial T)$ when $\rho$ depends on time only through $T$, Eq.~\eqref{coneq} implies
\begin{equation}
    \frac{dT}{dt}
    = \frac{-3H(\rho+P)+\delta\rho/\delta t}{\partial\rho/\partial T}.
    \label{eq:temp_t}
\end{equation}

In the remainder of this section we outline our solution strategy for neutrino decoupling. We follow the approximations of Refs.~\cite{Escudero:2020dfa,Escudero:2018mvt} (see also Ref.~\cite{Escudero:2025kej}) as implemented in the public \texttt{NUDEC\_BSM} code\footnote{\url{https://github.com/MiguelEA/nudec_BSM}}, which we modify to allow for a varying $m_e$. Specifically, we assume:
\begin{itemize}
    \item \emph{Thermal neutrino spectra:} neutrinos are described by Fermi--Dirac distributions with a (time-dependent) temperature. Non-thermal spectral distortions in the Standard Model affect the neutrino energy density at the $\sim 1\%$ level~\cite{Bennett:2020zkv}; given the level of precision targeted here, we neglect these distortions.
    \item \emph{No neutrino oscillations:} we neglect flavor oscillations during decoupling. Their impact on \Neff\ in the Standard Model is small (a few $\times 10^{-4}$) and well below the sensitivity expected for the effects considered here. We therefore evolve separate temperatures for $\nu_e$ and for $\nu_{\mu,\tau}$, capturing the leading flavor dependence from charged-current interactions.
    \item \emph{Vanishing chemical potentials:} we set $\mu_\gamma=\mu_e=\mu_{\nu_\alpha}=0$. The electron chemical potential is suppressed by the small baryon-to-photon ratio, and constraints on lepton asymmetries motivate negligible neutrino chemical potentials in the temperature range of interest (see the discussion in Appendix A.3 of Ref.~\cite{Escudero:2020dfa}).
\end{itemize}
Under these assumptions, for ultrarelativistic neutrinos one has $\rho_{\nu_{\alpha}} = g_{\nu_{\alpha}}\frac{7}{8}\frac{\pi^2}{30} T_{\nu_{\alpha}}^4$ and $P_{\nu_{\alpha}}=\rho_{\nu_{\alpha}}/3$.

The neutrino collision term, $\delta\rho_{\nu_{\alpha}}/\delta t$, must include interactions between neutrinos and the electromagnetic plasma, as well as neutrino self-interactions. Following Ref.~\cite{Escudero:2020dfa}, the relevant processes are:
\begin{itemize}
\item $e^+e^- \leftrightarrow \nu_{\alpha}\bar{\nu}_{\alpha}$,
\item $e^{\pm}\nu_{\alpha}\leftrightarrow e^{\pm}\nu_{\alpha}$,
\item $e^{\pm}\bar{\nu}_{\alpha} \leftrightarrow e^{\pm}\bar{\nu}_{\alpha}$,
\item $\nu_{\alpha}\nu_{\beta}\leftrightarrow \nu_{\alpha}\nu_{\beta}$,
\item $\nu_{\alpha}\bar{\nu}_{\beta}\leftrightarrow \nu_{\alpha}\bar{\nu}_{\beta}$,
\item $\nu_{\alpha}\bar{\nu}_{\alpha}\leftrightarrow \bar{\nu}_{\beta}\nu_{\beta}$.
\end{itemize}
Within the approximations above, the energy-transfer rates can be written in a compact form in terms of $T_\gamma$ and $T_{\nu_\alpha}$:
\begin{align}
\frac{\delta\rho_{\nu_{e}}}{\delta t}\Big|^{FD}_{SM} &= \frac{G_F^2}{\pi^5}\Big[4(g_{eL}^2+g_{eR}^2)\,F(T_{\gamma},T_{\nu_e})+2F(T_{\nu_\mu},T_{\nu_e})\Big],\label{deltarhonue}\\
\frac{\delta\rho_{\nu_{\mu}}}{\delta t}\Big|^{FD}_{SM} &= \frac{G_F^2}{\pi^5}\Big[4(g_{\mu L}^2+g_{\mu R}^2)\,F(T_{\gamma},T_{\nu_\mu})-F(T_{\nu_\mu},T_{\nu_e})\Big], \label{deltarhonumu}
\end{align}
where $G_F$ is the Fermi constant~\cite{ParticleDataGroup:2024cfk}, and $g_{eL}=0.727$, $g_{eR}=0.233$, $g_{\mu L}=-0.273$, $g_{\mu R}=0.233$~\cite{Escudero:2020dfa}. The function $F$ is
\begin{equation}
F(T_1,T_2) = 32 f_a^{FD}(T_1^9-T_2^9)+56 f_s^{FD}T_1^4 T_2^4(T_1-T_2),
\end{equation}
with $f_a^{FD}=0.884$ and $f_s^{FD}=0.829$, which account for Pauli-blocking effects in the annihilation and scattering rates, respectively~\cite{Escudero:2020dfa}. The expressions above are obtained in the $m_e\to 0$ limit of the interaction rates. While the electron-mass dependence is subdominant, we incorporate it by interpolating over the exact numerically precomputed rates including $m_e$, as provided in Ref.~\cite{Escudero:2020dfa}.

Equation~\eqref{eq:temp_t} then implies, for each neutrino flavor,
\begin{align}
    \frac{dT_{\nu_{\alpha}}}{dt}
    &= \frac{-3H(\rho_{\nu_\alpha}+P_{\nu_\alpha})+\delta\rho_{\nu_\alpha}/\delta t}{\partial\rho_{\nu_\alpha}/\partial T_{\nu_\alpha}}
    \notag\\
    &= -H\,T_{\nu_{\alpha}}+\frac{\delta\rho_{\nu_{\alpha}}/\delta t}{\partial\rho_{\nu_{\alpha}}/\partial T_{\nu_{\alpha}}}.
    \label{eq:nuT}
\end{align}
It is convenient to define the comoving neutrino temperature $z_{\nu_{\alpha}}\equiv T_{\nu_{\alpha}}a$ (while keeping $z\equiv T_\gamma a$ for photons) and recast the evolution in terms of $x'$.
Using $H=\dot a/a=\dot x'/x'$, $\bar\rho_{\nu_{\alpha}}=a^4\rho_{\nu_{\alpha}}=g_{\nu_{\alpha}}\frac{7}{8}\frac{\pi^2}{30}z_{\nu_{\alpha}}^4$, and $\bar H^2\equiv a^4H^2=\frac{8\pi}{3}G\bar\rho$, we obtain
\begin{equation}
    \frac{dz_{\nu_{\alpha}}}{dx'} = z_{\nu_{\alpha}}\,\frac{x'^5}{m_0^6}\,
    \frac{\delta\rho_{\nu_{\alpha}}/\delta t}{4\,\bar{\rho}_{\nu_{\alpha}}\,\bar{H}}.
    \label{eq:dz_neutrino}
\end{equation}
Here $\delta\rho_{\nu_{\alpha}}/\delta t$ is given by Eqs.~\eqref{deltarhonue} and~\eqref{deltarhonumu}, with Eq.~\eqref{deltarhonumu} applying also to $\nu_\tau$. Equations~\eqref{eq:dzodxp} and~\eqref{eq:dz_neutrino} fully determine the evolution of the photon and neutrino temperatures through neutrino decoupling and $e^\pm$ annihilation.

\subsection{\label{sec:calc}Numerical results}
We now solve the coupled system described above to obtain the evolution of the plasma temperatures. We implement the numerical integration in \texttt{Python}, starting from the public \texttt{NUDEC\_BSM} code~\cite{Escudero:2020dfa,Escudero:2018mvt} (see also Ref.~\cite{Escudero:2025kej}) and modifying it to allow for a varying $m_e$. The integration focuses on Eqs.~\eqref{eq:dzodxp} and~\eqref{eq:dz_neutrino}. We integrate over $x'$ corresponding to the standard range in $x=m_e a$, namely $x\in[0.01,100]$, which translates into
\begin{equation}
x'_{\min}=0.01\,\frac{m_0}{m_e},\qquad x'_{\max}=100\,\frac{m_0}{m_e}.
\end{equation}
We fix $m_0=m_{e,0}=0.511~\MeV$ and treat $m_e$ as a free parameter. In this part of the analysis we evolve $T_{\nu_e}\neq T_{\nu_\mu}$ (with $T_{\nu_\tau}=T_{\nu_\mu}$).

\begin{figure}[t]
    \centering
    \includegraphics[width=0.5\textwidth]{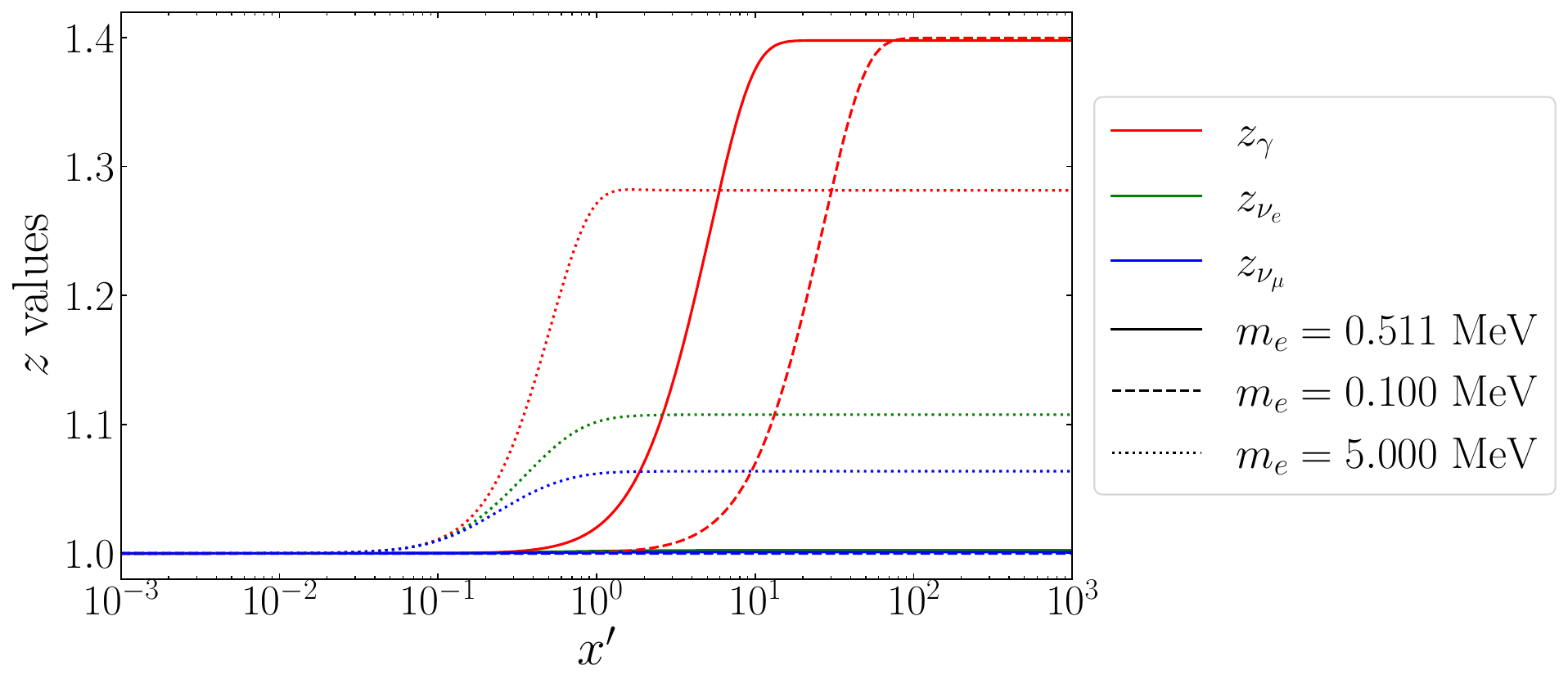}
    \caption{Evolution of the comoving temperatures $z_\gamma$, $z_{\nu_e}$, and $z_{\nu_\mu}$ as functions of $x'$.}
    \label{fig:z}
\end{figure}

Figure~\ref{fig:z} shows the evolution of $z_\gamma$, $z_{\nu_e}$, and $z_{\nu_\mu}$ as functions of $x'$ for three illustrative values of $m_e$: the standard $0.511~\MeV$ (solid), a smaller value $0.1~\MeV$ (dashed), and a larger value $5~\MeV$ (dotted). In the standard case, $e^\pm$ annihilation occurs when neutrinos are already close to decoupling, so only a small fraction of the released entropy is transferred to neutrinos. As a result, $z_\gamma$ asymptotes to a value close to the instantaneous-decoupling limit, while $z_{\nu_e}$ and $z_{\nu_\mu}$ remain only slightly above unity. For larger $m_e$ (e.g.\ $5~\MeV$), $e^\pm$ annihilation occurs earlier, while neutrinos are still more efficiently coupled to the plasma; neutrinos then absorb a larger share of the entropy release, reducing the photon heating and increasing $z_{\nu_e}$ and $z_{\nu_\mu}$. For smaller $m_e$ (e.g.\ $0.1~\MeV$), annihilation occurs later, when neutrinos are more completely decoupled, and the evolution approaches the instantaneous-decoupling behavior with $z_\gamma\simeq 1.4$ (apart from FTQED corrections).

\begin{figure}[t]
    \centering
    \includegraphics[width=0.5\textwidth]{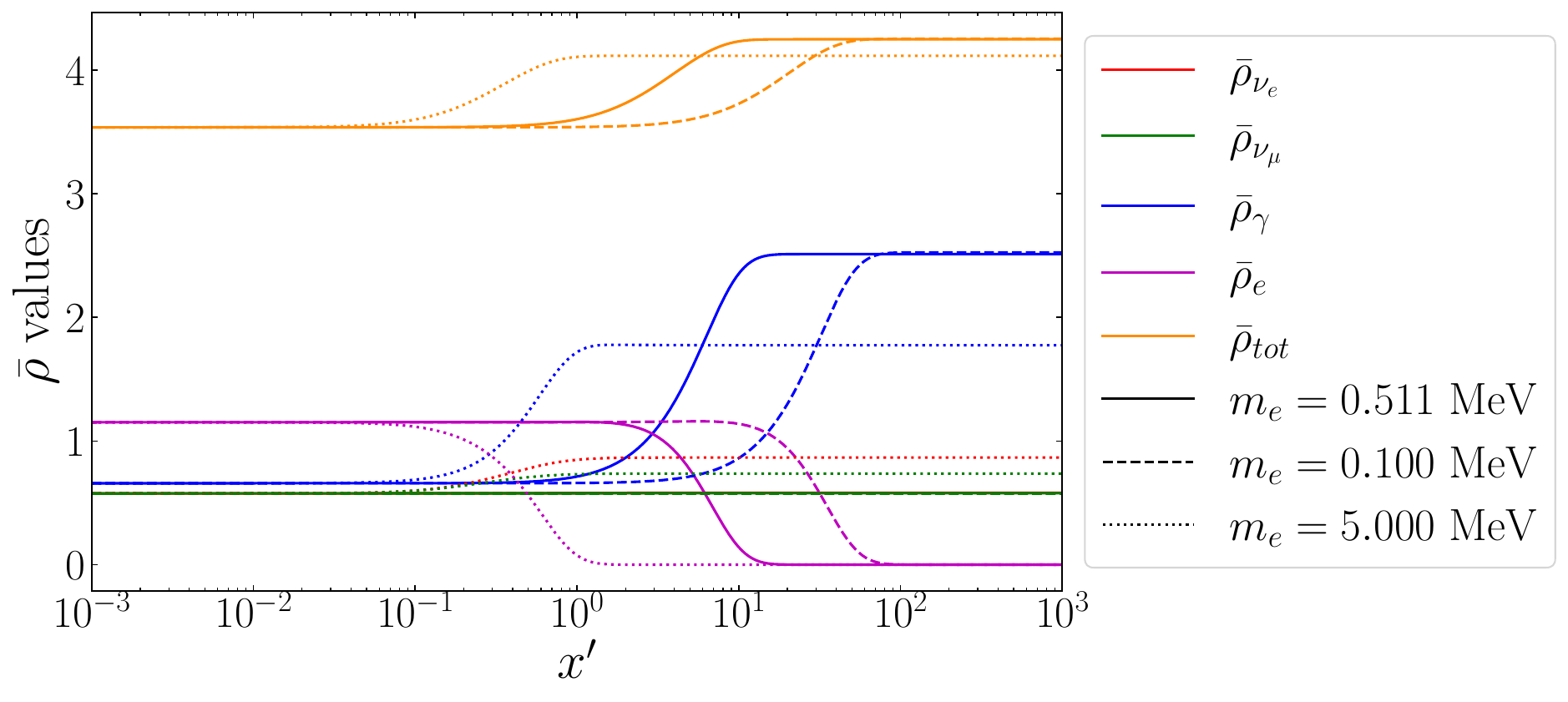}
    \caption{Evolution of the comoving energy density $\bar{\rho}$ as a function of $x'$.}
    \label{fig:rho}
\end{figure}

A consistent picture emerges when considering the comoving energy densities $\bar\rho_i$ as functions of $x'$, shown in Fig.~\ref{fig:rho}. If $e^\pm$ are heavier (here $m_e=5~\MeV$), their energy density becomes Boltzmann suppressed at earlier times than in the standard case ($m_e=0.511~\MeV$) or in the lighter-mass example ($m_e=0.1~\MeV$). When annihilation occurs while neutrinos are still coupled, part of the $e^\pm$ entropy is shared with the neutrino sector, and both $\bar\rho_\gamma$ and $\bar\rho_\nu$ increase accordingly. Conversely, if annihilation happens after neutrino decoupling, the dominant increase is in the photon energy density.

\begin{figure}[t]
    \centering
    \includegraphics[width=0.5\textwidth]{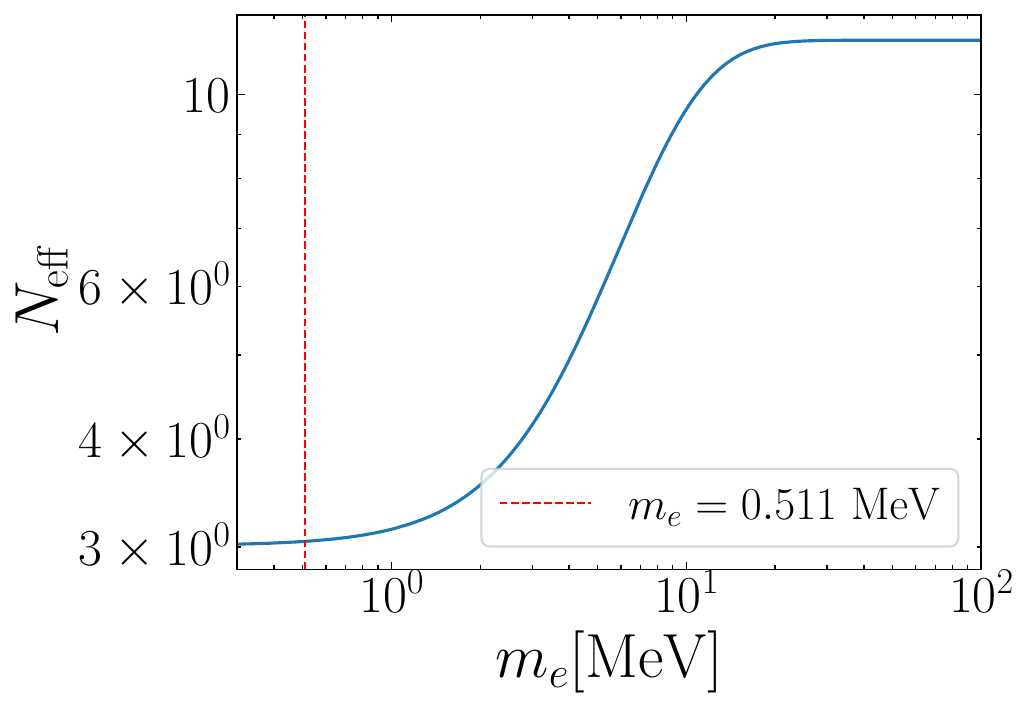}
    \caption{Evolution of \Neff\ as a function of $m_e$.}
    \label{fig:NEFF_NUOVO}
\end{figure}

Finally, we summarize the impact of a varying electron mass in terms of \Neff. Figure~\ref{fig:NEFF_NUOVO} shows \Neff\ as a function of $m_e$ obtained from the numerical solution of Eqs.~\eqref{eq:dzodxp} and~\eqref{eq:dz_neutrino}. For $m_e=0.511~\MeV$ the code reproduces the Standard Model prediction (within rounding, consistent with Ref.~\cite{Escudero:2020dfa}). As $m_e$ increases, \Neff\ grows and approaches the asymptotic value $\Neff \to 3(11/4)^{4/3}\simeq 11.56$, corresponding to the limiting case in which neutrinos remain tightly coupled to the electromagnetic plasma through $e^\pm$ annihilation. In this regime, neutrinos share the entropy released by $e^\pm$ annihilation, so the photon heating is reduced and the neutrino energy density is enhanced.

\section{\label{sec:bbn}Big Bang Nucleosynthesis}
Having established how a nonstandard electron mass $m_e$ affects neutrino decoupling, we now turn to the subsequent stage of cosmic evolution: Big Bang Nucleosynthesis (BBN). BBN provides one of the most robust probes of the early Universe, offering precise predictions for the primordial abundances of light nuclei synthesized from neutrons and protons.

In minimal (standard) BBN, the primordial yields are primarily determined by the baryon-to-photon ratio,
\begin{equation}
\eta_B \equiv \frac{n_B}{n_\gamma},\qquad \eta_{10}\equiv 10^{10}\eta_B,
\end{equation}
once Standard Model microphysics and the cosmic expansion history are specified. The parameter $\eta_B$ can be inferred independently from precision measurements of CMB temperature anisotropies, via the baryon density that controls the acoustic peak structure. This enables a comparison of the same quantity measured at two widely separated epochs. In standard cosmology, $\eta_B$ is conserved between BBN and photon decoupling, so the CMB-inferred value can be used as an input for BBN predictions and confronted with the observed primordial abundances. Using CMB fits, Ref.~\cite{Yeh2022} finds $\eta_{10}=6.040\pm 0.118$, which reproduces the observed primordial abundances well, with the notable exception of lithium. This longstanding discrepancy---the ``lithium problem''~\cite{Fields2020,Fields2011}---will be discussed further below.

Light-element synthesis is sensitive to the physical conditions in the radiation-dominated Universe at temperatures $T\sim (0.1 - 0.01)~\MeV$, corresponding to timescales from $\mathcal{O}(10^2)\,\mathrm{s}$ to $\mathcal{O}(10^4)\,\mathrm{s}$. At higher temperatures, weak interactions maintain chemical equilibrium between neutrons and protons, yielding
\begin{equation}
\frac{n}{p} = e^{-Q/T},
\end{equation}
where $Q=1.293~\MeV$ is the neutron--proton mass difference. As the temperature decreases, the $n\leftrightarrow p$ interconversion rate per nucleon falls faster than the Hubble rate, and chemical equilibrium is lost (``freeze-out'') at $T_{\rm fr}\sim \mathcal{O}(1)\,\MeV$, when $n/p \simeq e^{-Q/T_{\rm fr}}\approx 1/6$. The freeze-out ratio is sensitive to both microphysics and the expansion rate: $Q$ is set by strong and electromagnetic effects, while $T_{\rm fr}$ depends on weak interaction rates and gravity. After freeze-out, free neutrons undergo $\beta$-decay while the inverse process becomes inefficient, reducing the ratio to $n/p\sim 1/7$ by the time nuclear reactions begin.

Deuterium is the first nuclear bound state to form efficiently. Despite its binding energy $\Delta_D\simeq 2.22~\MeV$, deuterium production is delayed by photodissociation from the high-energy tail of the photon distribution until the photon temperature drops to $T\sim 0.07~\MeV$ (equivalently $t\sim 180~\mathrm{s}$). Observationally, deuterium is widely regarded as of primordial (pregalactic) origin~\cite{Epstein1976,Coc_2017,Kislitsyn_2024,schoneberg20242024bbnbaryonabundance}, since it is weakly bound and is efficiently destroyed when cycled through stars. Consequently, the primordial D abundance is measured in distant, chemically unprocessed environments where stellar processing is minimal. The PDG compilation~\cite{ParticleDataGroup:2024cfk} quotes
\begin{equation}
\left.\frac{\mathrm{D}}{\mathrm{H}}\right|_p \times 10^{6} = 25.47 \pm 0.29,
\end{equation}
a weighted mean of the measurements reported in Refs.~\cite{Cooke2014,Cooke2016,RiemerSorensen2015,Balashev2016,RiemerSorensen2017,Zavarygin2018,Cooke2018}.

Once deuterium can survive, helium-3 and tritium are produced and rapidly processed into helium-4, the most abundant and stable BBN product. Since the majority of neutrons available at the onset of nucleosynthesis end up bound in \ce{^4He}, the primordial helium mass fraction can be estimated as
\begin{equation}
    Y_P \equiv \frac{\rho(\ce{^4He})}{\rho_b}
    \simeq \frac{2(n/p)}{1+(n/p)}
    \simeq 0.25,
    \label{eq:helium4}
\end{equation}
where $\rho_b$ is the total baryonic mass density. The primordial \ce{^4He} abundance is inferred from recombination emission lines of \ce{He} and \ce{H} in metal-poor extragalactic H\,\textsc{ii} regions. Because \ce{^4He} is also synthesized in stars, the observed helium abundance must be extrapolated to zero metallicity, exploiting the expected correlation between helium and metal production. The PDG~\cite{ParticleDataGroup:2024cfk} quotes $Y_P = 0.245 \pm 0.003$, reported as a weighted mean of Refs.~\cite{Aver2021,Valerdi2019,Fernandez2019,Kurichin2021,Hsyu2020,Valerdi2021,Aver12021}.
During the final stages of manuscript preparation, a new helium determination appeared as a preprint~\cite{Aver:2026dxv} (see also~\cite{Skillman:2026ltj,Yeh:2026pil}), reporting $Y_P = 0.2458 \pm 0.0013$. This central value, obtained from the Large Binocular Telescope (LBT) observations of 54 metal-poor  H\,\textsc{ii} regions, is consistent with earlier estimates but with improved precision. In our analysis, we consider both the LBT and the PDG measurements and derive the bounds on the electron mass for both cases.

The third stable element produced in appreciable amounts during BBN is \ce{^7Li}. Its primordial abundance is commonly inferred from old, metal-poor halo stars, where \ce{^7Li} is observed to be approximately constant over a range of metallicities (the Spite plateau). It is useful to quantify metallicity through
\begin{equation}
[\mathrm{Fe}/\mathrm{H}] = \log_{10} \left( \frac{N_{\mathrm{Fe}}}{N_{\mathrm{H}}} \right)_{\text{star}}
- \log_{10} \left( \frac{N_{\mathrm{Fe}}}{N_{\mathrm{H}}} \right)_{\odot},
\end{equation}
which measures the iron abundance relative to the Sun. Observations indicate a marked drop and increased scatter in $\ce{^7Li}/\mathrm{H}$ at very low metallicity, particularly for $[\mathrm{Fe}/\mathrm{H}]<-3.0$ and down to $[\mathrm{Fe}/\mathrm{H}]\simeq -4.5$, complicating a robust extrapolation to zero metallicity. The PDG~\cite{ParticleDataGroup:2024cfk} quotes
\begin{equation}
\left.\frac{\ce{^7Li}}{\mathrm{H}}\right|_p = (1.6 \pm 0.3)\times 10^{-10},
\end{equation}
based on stars with metaliticity in the range $-2.8<[\mathrm{Fe}/\mathrm{H}]<-1.5$. A well-known tension persists between this value and the standard BBN prediction, which for a baryon density consistent with deuterium and CMB data is $\ce{^7Li/H}\sim 5.4\times 10^{-10}$~\cite{Fields2020,Fields2011}. This factor of about 3 of discrepancy may reflect astrophysical systematics in stellar lithium abundances and/or their interpretation, missing or underestimated nuclear inputs to BBN, or physics beyond the minimal scenario.

In the following, we compute BBN predictions using the \texttt{PRyMordial} code~\cite{Burns:2023sgx}\footnote{\url{https://github.com/vallima/PRyMordial/tree/main}}, which evolves a nuclear reaction network across three matched regimes: (i) a high-temperature stage in which the evolution can be restricted to nucleons (starting at $T=\mathcal{O}(10)\,\MeV$ down to temperatures near neutrino decoupling), (ii) an intermediate stage from $\mathcal{O}(1)\,\MeV$ to $\mathcal{O}(0.1)\,\MeV$ in which photodissociation of bound states is important, and (iii) a low-temperature stage beginning at $\mathcal{O}(0.1)\,\MeV$ where the full set of nuclear species is tracked down to $T=\mathcal{O}(1)\,\keV$. The solution in each regime provides initial conditions for the subsequent stage.

\begin{figure}[t]
    \centering
    \includegraphics[width=0.5\textwidth]{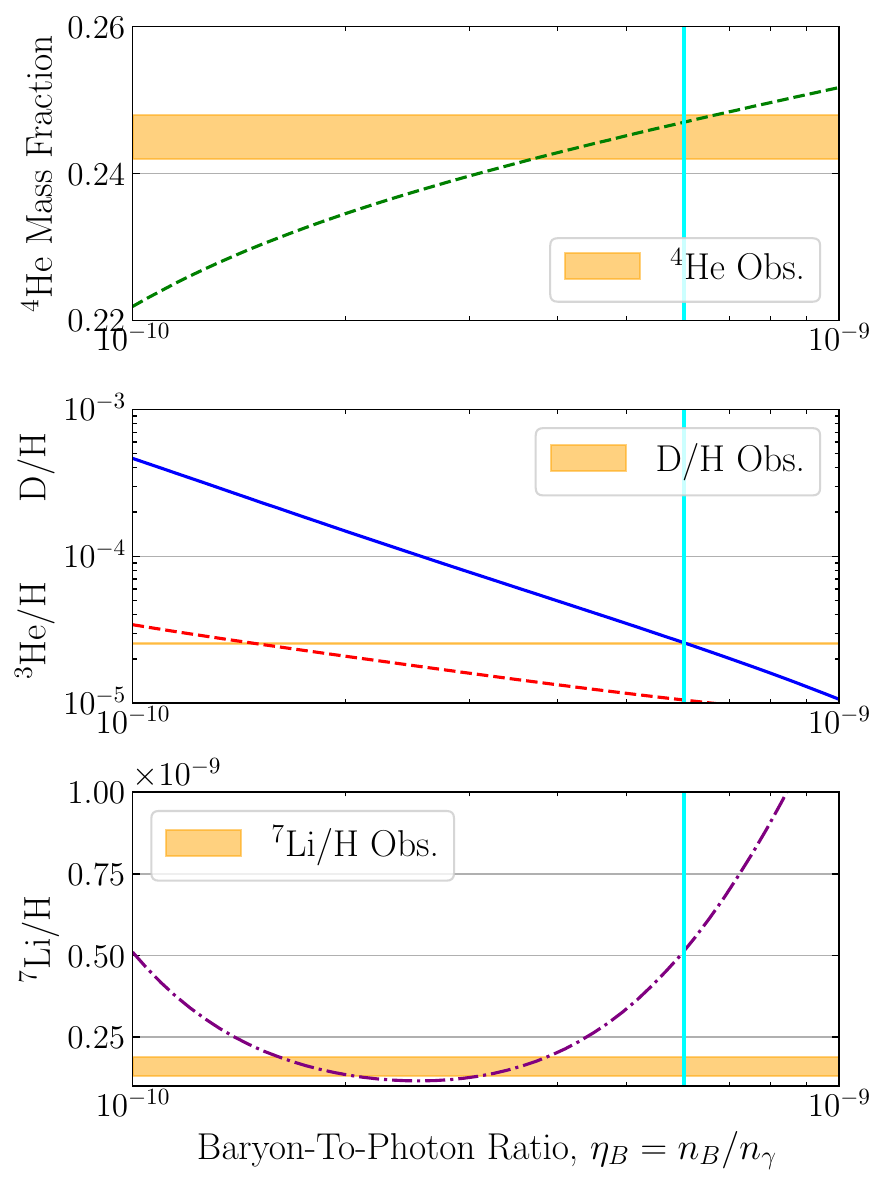}
    \caption{Primordial abundances of \ce{^4He}, \ce{D}, \ce{^3He}, and \ce{^7Li} as predicted by \texttt{PRyMordial} within the Standard Model (NACRE II compilation package used for the key processes), as a function of the cosmic baryon density. The cyan vertical band indicates the CMB constraint $\eta_{10} = 6.040 \pm 0.118$~\cite{Yeh2022}.}
    \label{fig:pdg}
\end{figure}

Figure~\ref{fig:pdg} shows the primordial abundances of \ce{^4He}, \ce{D}, \ce{^3He}, and \ce{^7Li} predicted by \texttt{PRyMordial} within the Standard Model as functions of the baryon density. The observed light-element abundances (yellow) and the CMB constraint on the baryon-to-photon ratio (cyan) are taken from Fig.~24.1 of the PDG~\cite{ParticleDataGroup:2024cfk}. Overall, the agreement between standard BBN predictions and the inferred primordial abundances is striking (again with the exception of \ce{^7Li}), providing strong support for the hot Big Bang picture and the Standard Model description of the early Universe~\cite{Schramm_1998,Steigman_2007,Iocco_2009,Cyburt_2016,ParticleDataGroup:2024cfk}. The ``lithium problem'' is clearly visible in Fig.~\ref{fig:pdg}: the observed \ce{^7Li/H} corresponds to values of $\eta_{10}$ incompatible with those preferred by \ce{D/H} (and also with \ce{^4He}, albeit less constraining).

Up to this point we have assumed the standard electron mass, $m_e=0.511~\MeV$. Since the weak rates governing $n\leftrightarrow p$ conversion depend on $m_e$, varying $m_e$ modifies the freeze-out value of $n/p$ and thus the initial conditions for light-element synthesis, as we discuss next.

\section{\label{sec:analysis}BBN with a non-standard electron mass}
The cosmological implications of an electron mass different from its laboratory value have been explored in a variety of contexts, including Refs.~\cite{Toda:2024ncp,Planck2015fundamentalconstants,Chluba2016,Hart2017}, primarily through its impact on recombination (and hence on the sound horizon) and in connection with possible resolutions of the Hubble tension. In this work, instead, we focus on an earlier epoch---the first few minutes---when neutrino decoupling and BBN occur. The key point is that light-element abundances provide a sensitive probe of variations in $m_e$, because $m_e$ affects (i) the weak rates governing neutron--proton interconversion, (ii) the available phase space for neutron $\beta$-decay and therefore the neutron lifetime, and (iii) the expansion history through the thermodynamics of the electromagnetic plasma and the associated change in \Neff.

The weak rates depend on the kinematics and on the phase-space volume available to the final-state particles. Increasing $m_e$ raises the energy threshold for producing an electron in processes such as
$n \rightarrow p + e^- + \bar{\nu}_e$, and reduces the kinetic energy available to be shared among the decay products. Both effects decrease the phase-space volume and therefore suppress the decay rate.
In order to keep neutron decay possible, we restrict our analysis to the case $m_e<Q = 1.293$ MeV when BBN is involved. This will not affect our final bounds on $m_e$, since they will turn out to be very close to the standard value of the electron mass $m_e = 0.511$ MeV.

Using the general scaling
$\Gamma \propto \int |{\cal M}|^2 \times (\text{phase space})$,
a reduced phase-space volume implies a smaller decay rate and hence a longer neutron lifetime.
Following the notation of Ref.~\cite{burns2023}, we write
\begin{equation}
\tau_n^{-1}=\tilde{G}_F^{\,2}\,m_e^5\,\mathcal{F}_n,
\end{equation}
where $\mathcal{F}_n$ (defined precisely in Ref.~\cite{Cirigliano2023}) includes the zero-temperature phase-space factor and electroweak radiative corrections~\cite{Marciano_2006}, and
\begin{equation}
\tilde{G}_F \equiv G_F\,V_{\rm ud}\,\sqrt{\frac{1+3g_A^2}{2\pi^3}}\,,
\end{equation}
with $V_{\rm ud}$ related to the Cabibbo angle~\cite{utfitcollaboration2022newutfitanalysisunitarity} and $g_A$ the nucleon axial coupling~\cite{PhysRevD.88.073002}.
Since $\tau_n$ is a critical ingredient in setting the neutron-to-proton ratio at the onset of nucleosynthesis, even modest changes in $m_e$ can propagate into observable shifts in the predicted light-element abundances. For instance, if neutrons survive longer than in the standard case, the neutron abundance at the onset of nuclear reactions increases, favoring a larger \ce{^4He} mass fraction.

\begin{figure}[t]
    \centering
    \includegraphics[width=0.5\textwidth]{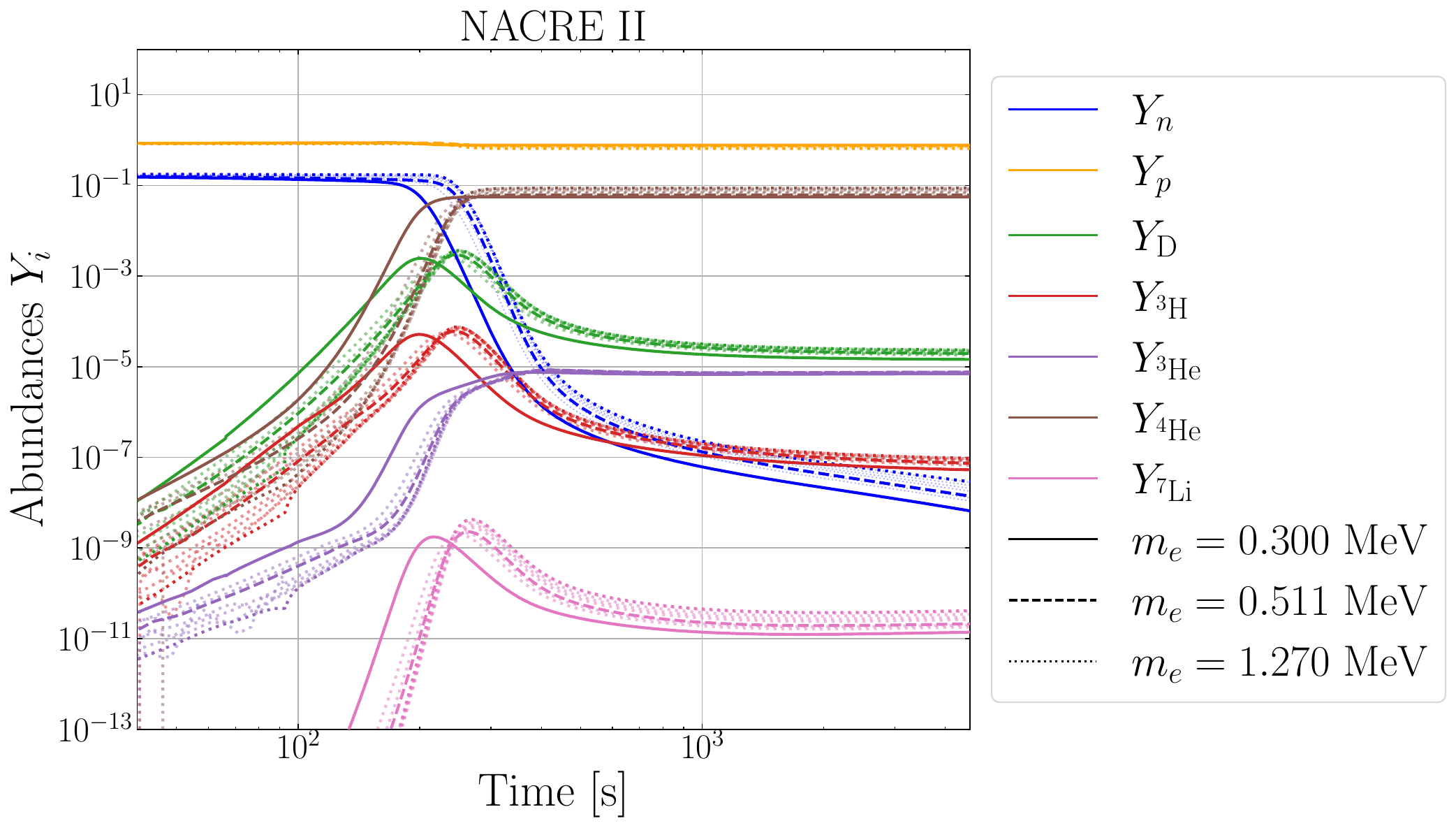}
    \includegraphics[width=0.5\textwidth]{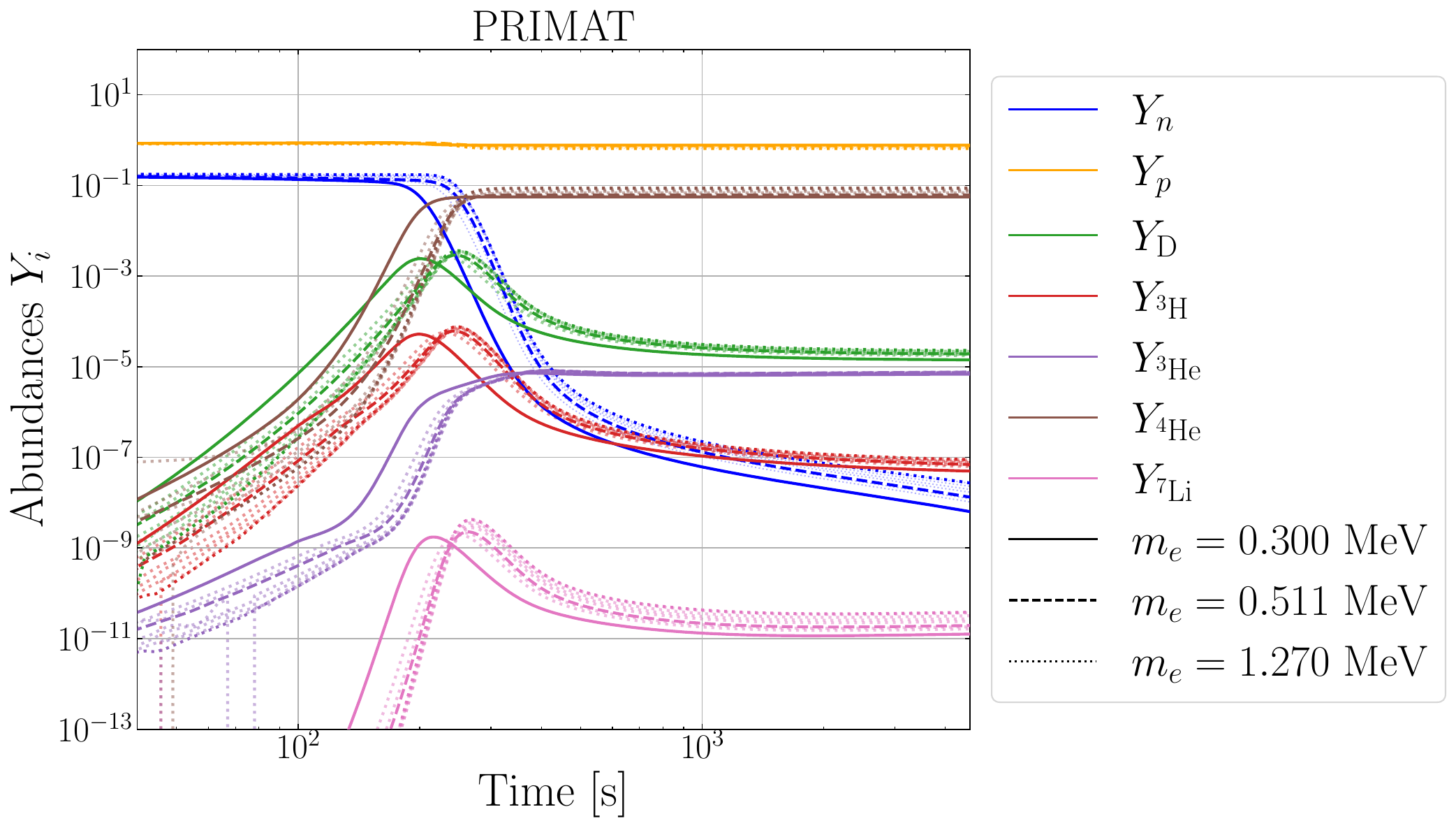}
    \caption{Time evolution of primordial abundances for different values of $m_e$, using the NACRE II (top panel) and PRIMAT (bottom panel) compilations.}
    \label{fig:me_abundance_comparison}
\end{figure}

We study these effects with a modified version of the \texttt{PRyMordial} code~\cite{Burns:2023sgx} that allows for variations in $m_e$ and retains the refinements beyond the Born approximation needed for accurate BBN predictions. In particular, the code includes:
\begin{itemize}
    \item QED radiative corrections (vacuum) to the $n \leftrightarrow p$ amplitudes at $\mathcal{O}(\alpha_{\rm em})$, including virtual and real-photon emission;
    \item finite nucleon-mass effects and weak magnetism, which induce relative shifts of order $\Delta\Gamma/\Gamma\sim~\mathcal{O}(10^{-2})$ in the weak rates;
    \item finite-temperature corrections required for sub-percent accuracy.
\end{itemize}
These ingredients are implemented in the public \texttt{PRyMordial} release. That code also allows one to treat the neutron lifetime as an external parameter (by setting \texttt{tau\_n\_flag=False}); here we instead regard $m_e$ as the fundamental input and let the corresponding weak rates (and hence the effective $\tau_n$) follow from the modified microphysics, rather than imposing the present-day neutron-lifetime measurement.
Regarding the values of $G_F, V_{\rm ud}, Q,g_A$ and other fundamental physics parameters, these are kept fixed at their standard values.
With these assumptions, \texttt{PRyMordial} predicts the time evolution of the light-element abundances $Y_i(t)$, shown in Fig.~\ref{fig:me_abundance_comparison}. The top and bottom panels differ only in the compilation adopted for thermonuclear reaction rates:
\begin{itemize}
    \item The top panel uses the NACRE II compilation~\cite{Xu2013NACREII}, which provides evaluated charged-particle reaction rates for targets with $A<16$ (based on a potential-model description of cross sections in the relevant energy range);
    \item The bottom panel uses the PRIMAT compilation~\cite{Pitrou_2018,Pitrou_2021}, an extensive catalogue (hundreds of reactions) with cross sections obtained from refined statistical analyses within $R$-matrix theory or computed with dedicated tools such as \texttt{TALYS}.
\end{itemize}
The two compilations are known to differ in specific channels~\cite{Pitrou_2018,Pitrou_2021,Xu2013NACREII}, notably in the deuterium prediction, which is systematically lower in PRIMAT. We nevertheless consider both, since NACRE II is a general-purpose compilation for a broad range of astrophysical applications, whereas PRIMAT is designed to maintain internal consistency within a complete BBN framework (thermodynamics, weak sector, and network), enabling a particularly direct comparison to the standard theoretical baseline.

In Fig.~\ref{fig:me_abundance_comparison}, the dashed curves correspond to the standard case $m_e=0.511~\mathrm{MeV}$. The solid and dash-dotted curves illustrate two representative nonstandard values, $m_e=0.300~\mathrm{MeV}$ and $m_e=1.270~\mathrm{MeV}$, while the lighter curves show intermediate masses used to map the overall trend. Qualitatively, increasing $m_e$ tends to (i) suppress the weak rates through reduced phase space and altered $e^\pm$ densities and (ii) increase the neutron lifetime, both of which favor a larger neutron abundance when nucleosynthesis begins. At the same time, changing $m_e$ also affects the expansion history through the plasma equation of state and through the associated shift in \Neff. The net result in our numerical solutions is that larger $m_e$ values yield a delayed onset of the reaction network in time, but with a higher available neutron fraction, leading to enhanced production of neutron-rich nuclei, most prominently \ce{^4He}. Conversely, smaller $m_e$ values tend to increase the weak rates and shorten $\tau_n$, reducing the neutron abundance at the onset of nucleosynthesis and shifting the network evolution accordingly. The proton abundance is comparatively less sensitive, since it is set primarily by baryon number conservation and by the much smaller neutron fraction.

\begin{figure}[t]
    \centering
    \includegraphics[width=0.5\textwidth]{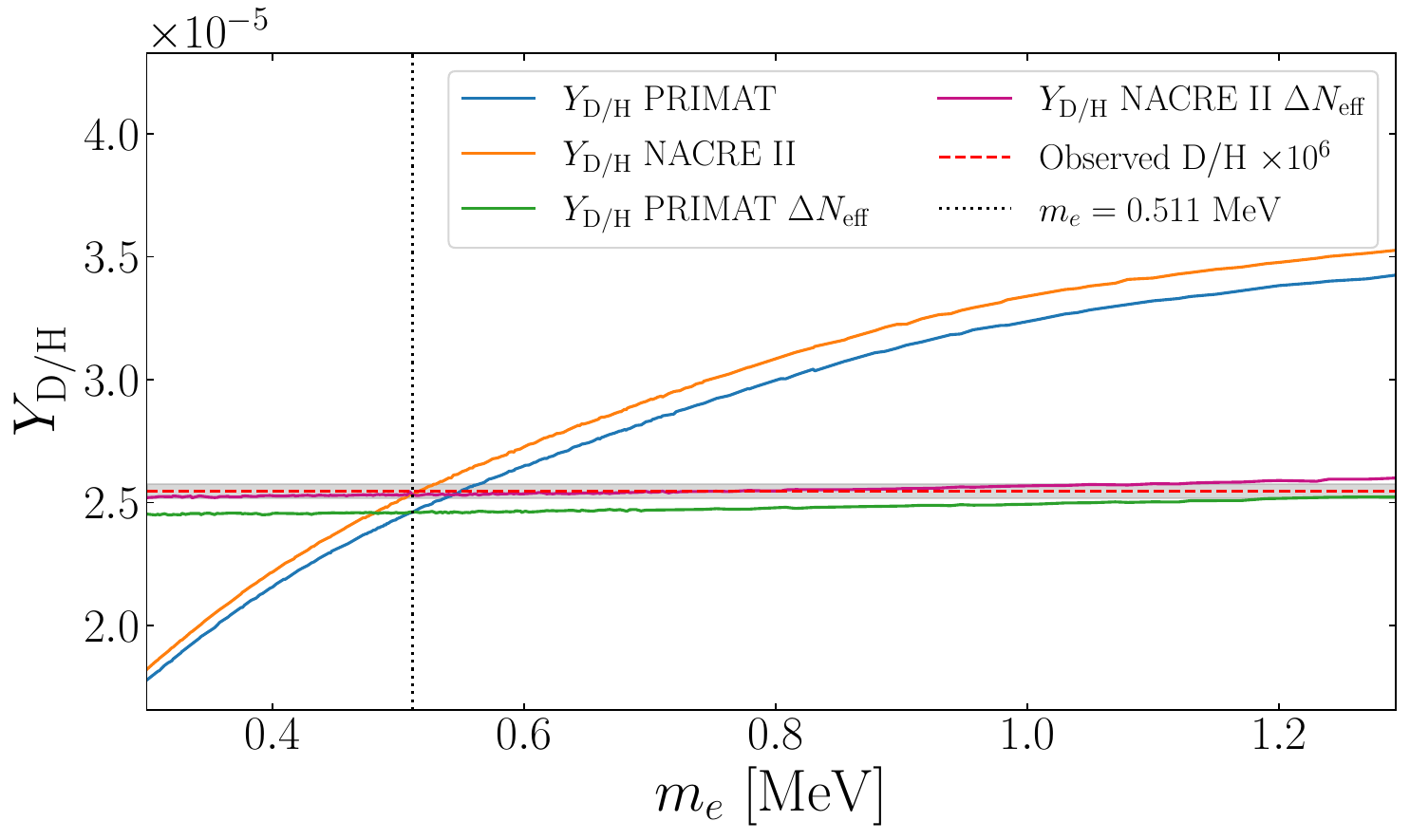}
    \caption{Deuterium abundance as a function of $m_e$.}
    \label{fig:deuterium_abundance}
\end{figure}

\begin{figure}[t]
    \centering
    \includegraphics[width=0.5\textwidth]{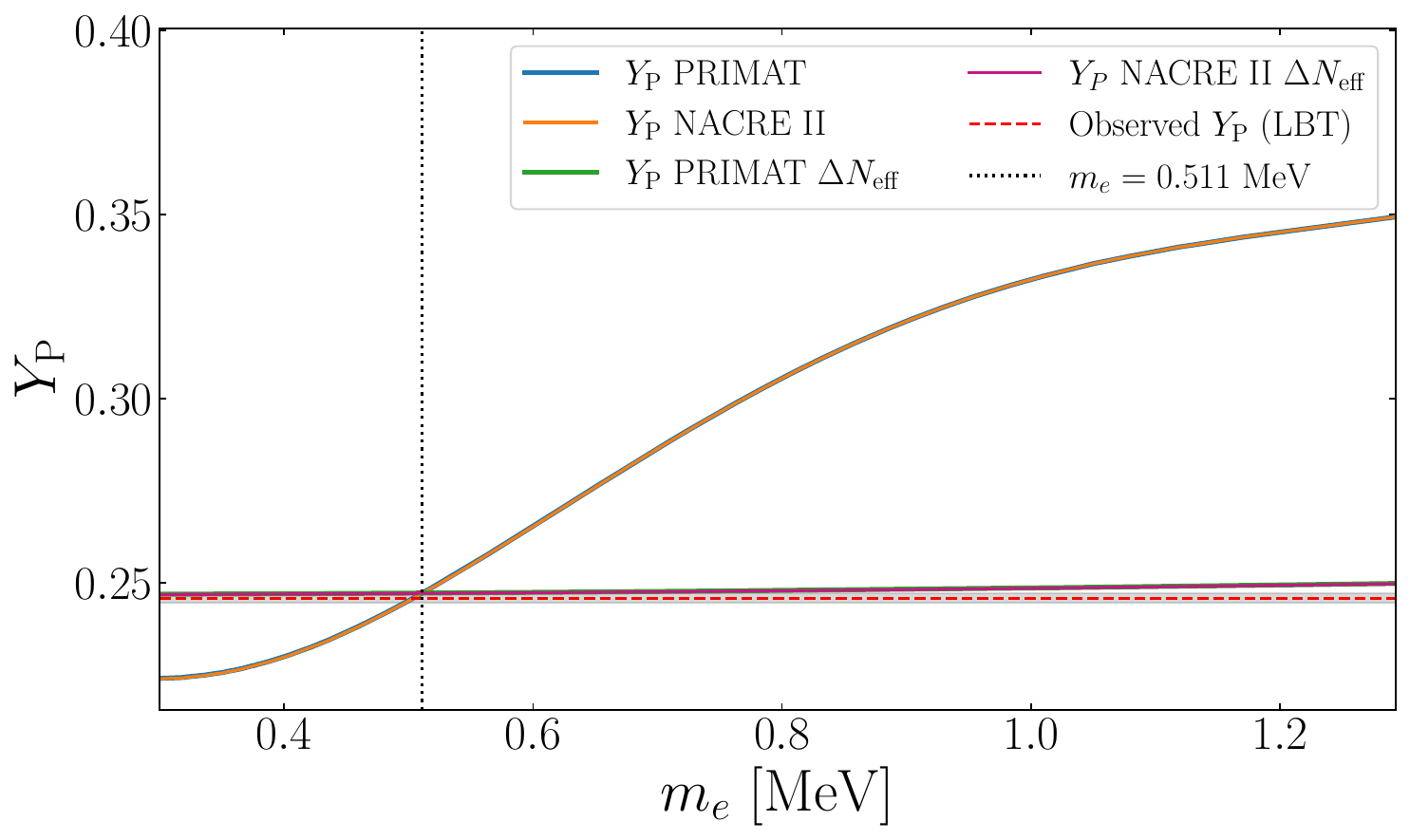}
    \caption{Helium-4 abundance as a function of $m_e$.}
    \label{fig:helium4_abundance}
\end{figure}

\begin{figure}[t]
    \centering
    \includegraphics[width=0.5\textwidth]{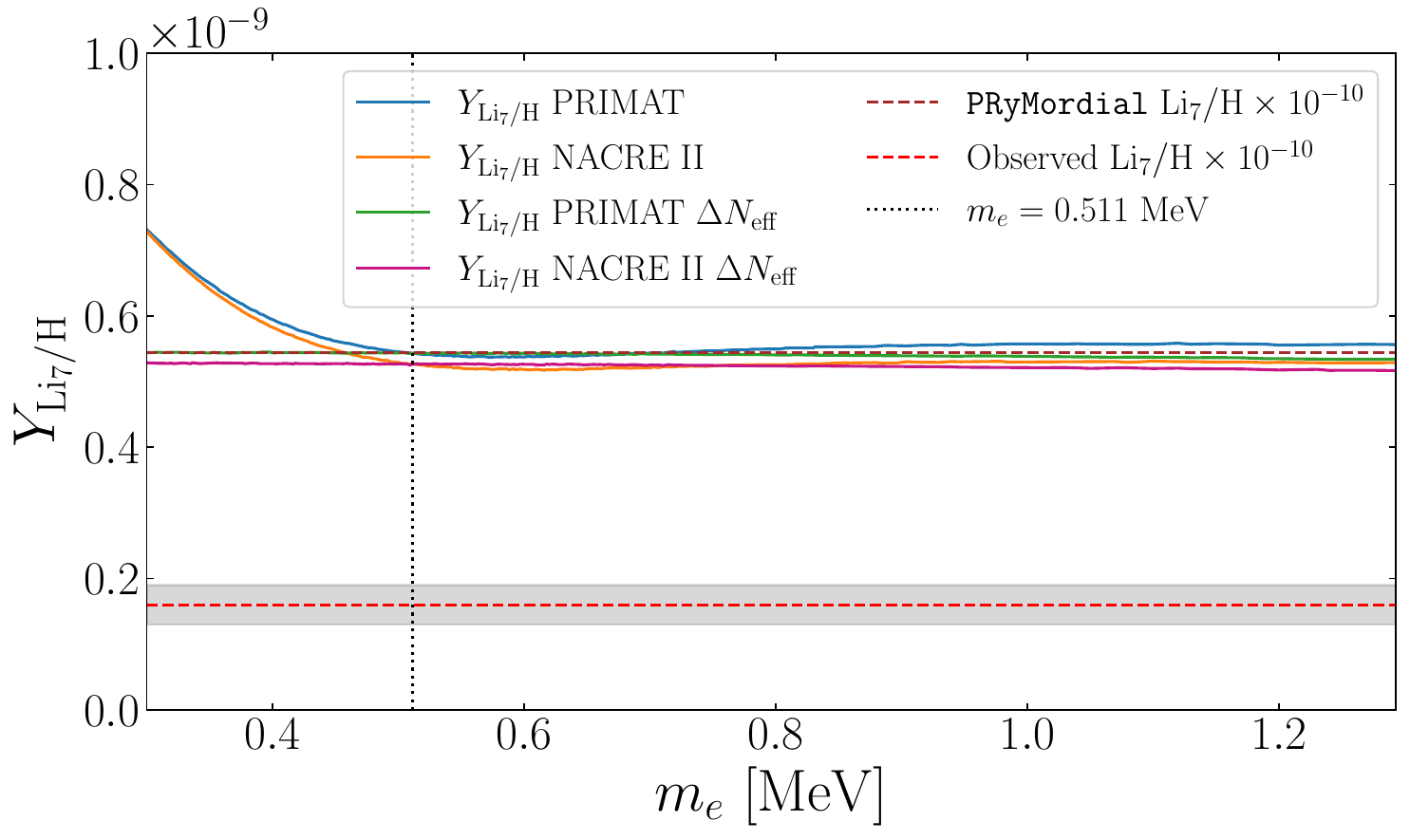}
    \caption{Lithium-7 abundance as a function of $m_e$.}
    \label{fig:lithium7_abundance}
\end{figure}

The same study can be summarized by considering the final abundances of deuterium, helium-4, and lithium-7 as functions of $m_e$ (Figs.~\ref{fig:deuterium_abundance}--\ref{fig:lithium7_abundance}).
For deuterium and helium-4 (Figs.~\ref{fig:deuterium_abundance} and~\ref{fig:helium4_abundance}), the full calculation shows an overall increase of the predicted abundances with $m_e$ across the range considered here. This trend reflects the combined microphysical impact of $m_e$ on the weak sector (including $\tau_n$) and the thermodynamic/expansion effects discussed above. The difference between NACRE II (orange) and PRIMAT (blue) is most visible for deuterium, with PRIMAT predicting a systematically lower value for the ratio between the deuterium and hydrogen abundances (D/H). This is consistent with known differences between the compilations: PRIMAT adopts larger low-energy astrophysical $S$ factors for relevant channels than NACRE II, leading to more efficient deuterium destruction during BBN. The same sensitivity is not present for \ce{^4He}. Once the neutron abundance at the onset of nucleosynthesis is fixed, nearly all surviving neutrons are incorporated into \ce{^4He}, so small variations in nuclear rates have little effect on $Y_\mathrm{P}$. As a result, the NACRE II and PRIMAT predictions for \ce{^4He} largely overlap in Fig.~\ref{fig:helium4_abundance}.

\begin{figure*}[t]
\centering
\includegraphics[width=0.49\textwidth]{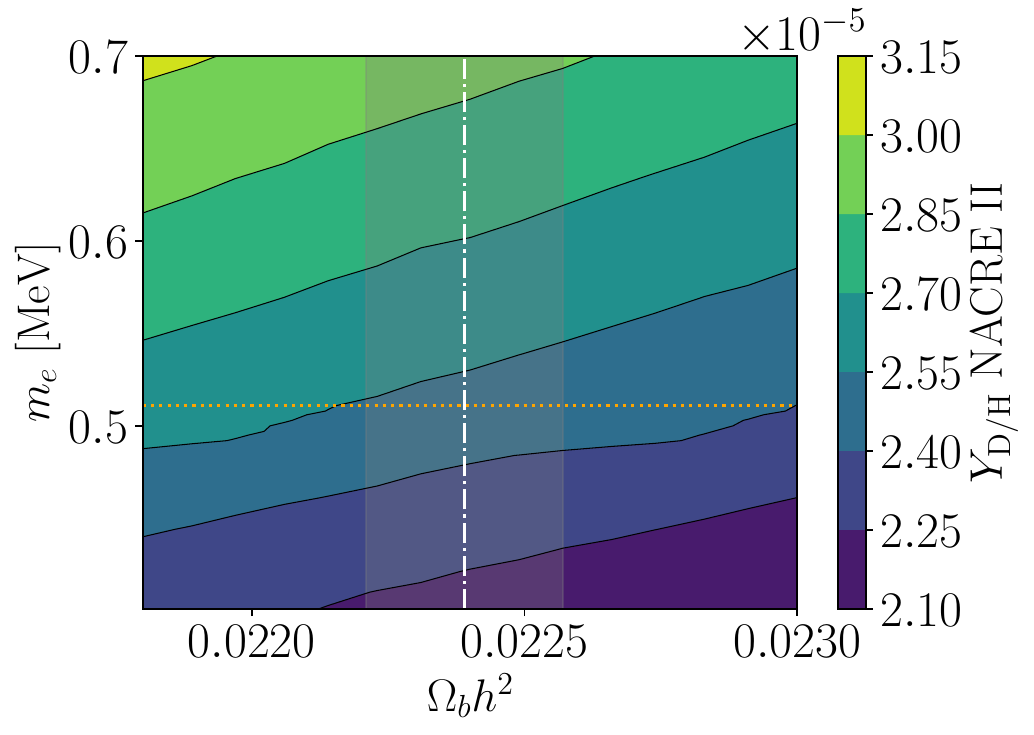}
\includegraphics[width=0.49\textwidth]{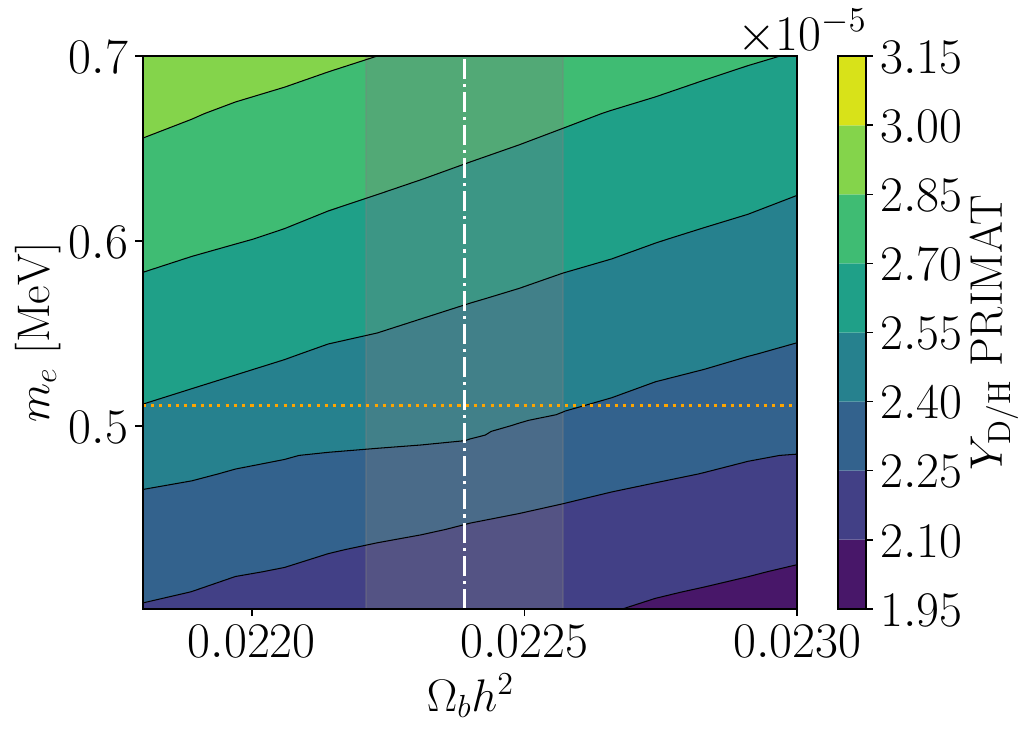}\\
\includegraphics[width=0.49\textwidth]{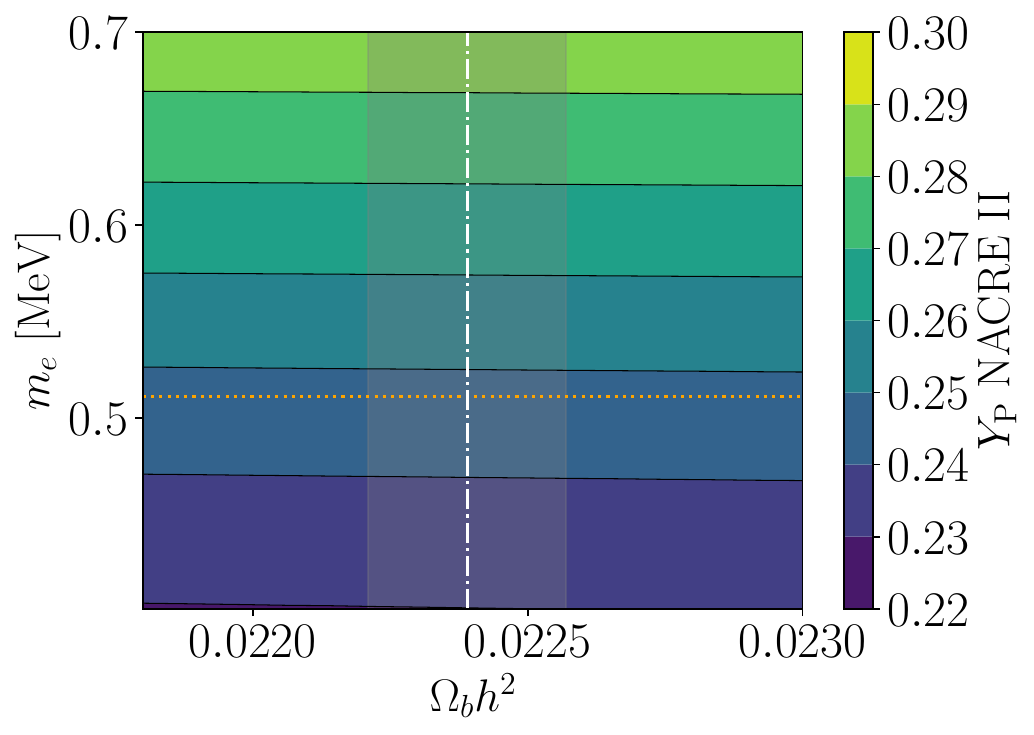}
\includegraphics[width=0.49\textwidth]{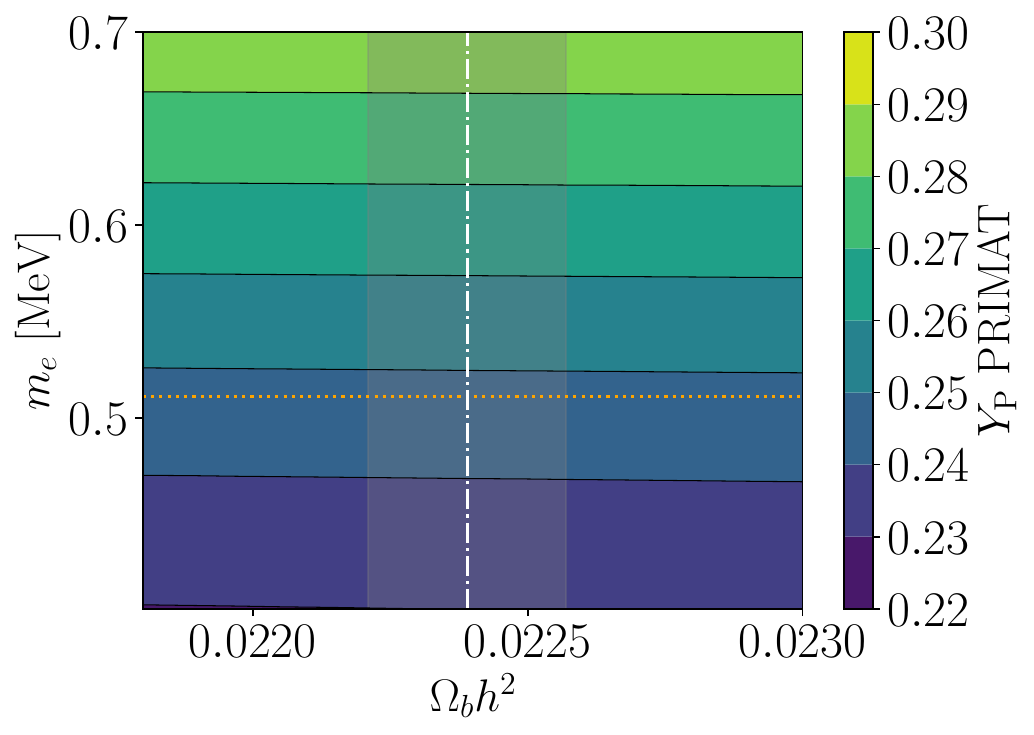}
\caption{Deuterium abundance (top row) and Helium-4 abundance (bottom row) as a function of $m_e$ and $\Omega_b h^2$ using the NACRE II (left column) or PRIMAT (right column) compilation. The orange horizontal line highlights $m_e=0.511~\mathrm{MeV}$, while the vertical white lines and overlayed band show the Planck \cite{Planck:2018vyg} constraint $\Omega_b h^2 = 0.02238\pm0.00019$.}
\label{fig:contours_bbn}
\end{figure*}

\looseness=-1
A similar information can be obtained by looking at the four panels of Fig.~\ref{fig:contours_bbn}. The contours represent lines of constant values for $Y_{\rm{D/H}}$ and $Y_{\rm{P}}$ using both NACRE II and PRIMAT, while the coloring of the contours shows how these quantities vary across the parameter space of $\Omega_b h^2$ and $m_e$. The panels show again that as $m_e$ increases, the abundances of both deuterium and helium-4 increase. Moreover, it is possible to focus on the dependence of these primordial nuclei on the baryon density parameter. Deuterium (top row of Fig.~\ref{fig:contours_bbn}) is more affected by changes in $\Omega_b h^2$: a higher baryon density leads to more deuterium production, but if it is too high, deuterium can be destroyed by nuclear reactions. In fact, the sloped contours show that a change in the baryon density could alter significantly the synthesis-destruction equilibrium of $Y_{\rm{D/H}}$ in the early Universe. On the other hand, the bottom row of Fig.~\ref{fig:contours_bbn} shows that the helium-4 contours are more horizontal, leading to the conclusion that $Y_{\rm{P}}$ is not altered by $\Omega_b h^2$ due to its limited relation with nuclear reactions.


To disentangle expansion effects from direct changes in the weak rates, we also study the impact of varying \Neff\ alone, keeping the weak rates fixed to their standard values. For this test, for reasons related to the treatment of neutrino temperatures in the \texttt{PRyMordial} code we assume a single neutrino temperature, $T_{\nu_e}=T_{\nu_{\mu,\tau}}$, and adopt the standard value $\Neff=3.046$ for $m_e=0.511~\mathrm{MeV}$ (consistent with the second-to-last line of Table~1 in Ref.~\cite{Escudero:2020dfa}). We then compute $\Neff(m_e)$ from Fig.~\ref{fig:NEFF_NUOVO} and propagate only
\begin{equation}
\Delta N_{\mathrm{eff}} \equiv N_{\mathrm{eff}}(m_e)-N_{\mathrm{eff}}(m_e=0.511~\mathrm{MeV})
\end{equation}
into \texttt{PRyMordial}, without altering the weak sector. Over the range $m_e\in[0.300,1.270]~\mathrm{MeV}$ we find $\Delta N_{\mathrm{eff}}~\in~[-0.0235,\,0.1872]$. The resulting abundance shifts (green and magenta curves in Figs.~\ref{fig:deuterium_abundance}--\ref{fig:lithium7_abundance}) are comparatively small: deuterium shows an almost flat dependence, while helium-4 exhibits only a mild increase with $m_e$ through the faster expansion associated with larger \Neff. This confirms that, in the parameter range considered here, the dominant impact of varying $m_e$ on BBN arises from the direct microphysical modifications of the weak rates (including $\tau_n$), with expansion effects playing a subleading role. 

The lithium-7 case is more intricate. We normalize to the standard \texttt{PRyMordial} prediction at $m_e=0.511~\mathrm{MeV}$, which is already in tension with observations. Since the ``lithium problem'' is often attributed to astrophysical systematics and/or nuclear inputs~\cite{Fields2020,Fields2011}, variations in fundamental parameters are not expected to provide a simple resolution. In our full BBN computation (orange/blue curves in Fig.~\ref{fig:lithium7_abundance}), low $m_e$ values lead to an enhanced \ce{^7Li} yield, followed by a decrease and eventual stabilization at larger $m_e$. This non-monotonic behavior reflects the interplay between the \ce{^7Be}/\ce{^7Li} production channels, the \ce{^3He} and tritium bottlenecks, and the available neutron abundance. In contrast, varying only \Neff\ (green/magenta) produces a nearly flat response, again highlighting that \ce{^7Li} is comparatively insensitive to the expansion history alone in this setup. Overall, the changes in the final abundances are driven primarily by the direct dependence of the weak sector and thermodynamics on $m_e$.

\section{\label{sec:res}Results and discussion}
In this section we derive constraints on the electron mass $m_e$ during the BBN epoch. We combine CMB bounds on \Neff\ with observations of primordial light-element abundances, as discussed in the previous sections. Specifically, we perform a $\chi^2$ analysis using deuterium, helium-4, and \Neff, considering both their individual contributions and their combined constraint.
Furthermore, in the global analysis we have added a scan on the $\Omega_b h^2$ parameter, in order to appropriately take into account the dependence of the deuterium abundance on this parameter. We do not include \ce{^7Li} in the fit: the well-known tension between the observed and standard-BBN predictions is likely dominated by astrophysical systematics and/or poorly understood nuclear inputs, so incorporating lithium alongside better-controlled observables could bias the inference.

For each observable we define
\begin{align}
\chi^2_i &=
\left(
\frac{X_i^{\rm th}(m_e) - X_i^{\rm obs}}
     {\sqrt{\sigma_{i,{\rm exp}}^2 + \sigma_{i,{\rm th}}^2}}
\right)^2,
\label{eq:chi2_i}
\\
\chi^2_{\rm tot} &= \sum_i \chi^2_i,
\label{eq:chi2_tot}
\end{align}
where $X_i \in \left\{ \left(\frac{\mathrm{D}}{\mathrm{H}}\right)_p,\, Y_{\rm P},\, N_{\rm eff},\, \Omega_b h^2 \right\}$, $\sigma_{i,\rm exp}$ is the quoted experimental uncertainties and $\sigma_{i,\rm the}$ the theoretical uncertainty. The observational inputs are taken from the PDG~\cite{ParticleDataGroup:2024cfk} for deuterium, from the LBT results in \cite{Aver:2026dxv} and the PDG~\cite{ParticleDataGroup:2024cfk} for helium 4,
while we take the values obtained from Planck~\cite{Planck:2018vyg} (full CMB plus lensing and BAO observations for the $\Lambda$CDM+\Neff+$Y_P$ model) for \Neff\ and $\Omega_{\rm b}h^2$. We therefore use:
\begin{eqnarray}
\left.\frac{\mathrm{D}}{\mathrm{H}}\right|_p \times 10^{6} &=& 25.47 \pm 0.29, \\
Y_{\rm{P}} &=& 0.2458 \pm 0.0013\;(\mbox{LBT}), \\
Y_{\rm{P}} &=& 0.245 \pm 0.003\,(\mbox{PDG}), \\
N_{\mathrm{eff}}^{\rm obs} &=& 2.99 \pm 0.17\,(\mbox{fixed }Y_{\rm{P}}), \\
N_{\mathrm{eff}}^{\rm obs} &=& 2.97 \pm 0.29\,(\mbox{free }Y_{\rm{P}}), \label{eq:Neff} \\
(\Omega_{\mathrm{b}}h^2)^{\rm obs} &=& 0.02238 \pm 0.00019\,(\mbox{free }Y_{\rm{P}}).
\end{eqnarray}

The theoretical predictions $X_i^{\rm th}(m_e)$ are obtained from the modified \texttt{PRyMordial} runs (for D/H and $Y_{\rm{P}}$), from the neutrino-decoupling calculation (for \Neff).
For $\Omega_b h^2$ we consider 15 values in a $3\sigma$ range around the central value defined above. For the BBN theory errors we adopt the rate-induced uncertainties quoted in Ref.~\cite{Barenboim:2025okj}, namely
\begin{align}
\sigma_{Y_P,{\rm th}} &= 1.1\times 10^{-4}\ \ (1.4\times 10^{-4}),\\
\sigma_{{\rm D/H},{\rm th}} &= 2.6\times 10^{-7}\ \ (1.0\times 10^{-6}),
\end{align}
for PRIMAT (NACRE II), respectively. We neglect the theoretical uncertainty on \Neff, which is expected to be much smaller than the current experimental error even in our simplified setup~\cite{Bennett:2020zkv,Froustey:2020mcq,Escudero:2020dfa}.

For each dataset we compute $\chi^2(m_e)$ and present results in terms of
\begin{equation}
\Delta\chi^2(m_e)\equiv \chi^2(m_e)-\chi^2_{\rm min},
\end{equation}
which aligns the best-fit point at $\Delta\chi^2=0$ and facilitates a comparison of the relative preference for nearby values of $m_e$. 
For a single parameter, $\Delta\chi^2=1$ and $\Delta\chi^2=4$ correspond to the $1\sigma$ ($68.27\%$) and $2\sigma$ ($95.45\%$) confidence intervals, respectively.

\begin{figure*}[t]
\centering
\includegraphics[width=0.49\textwidth]{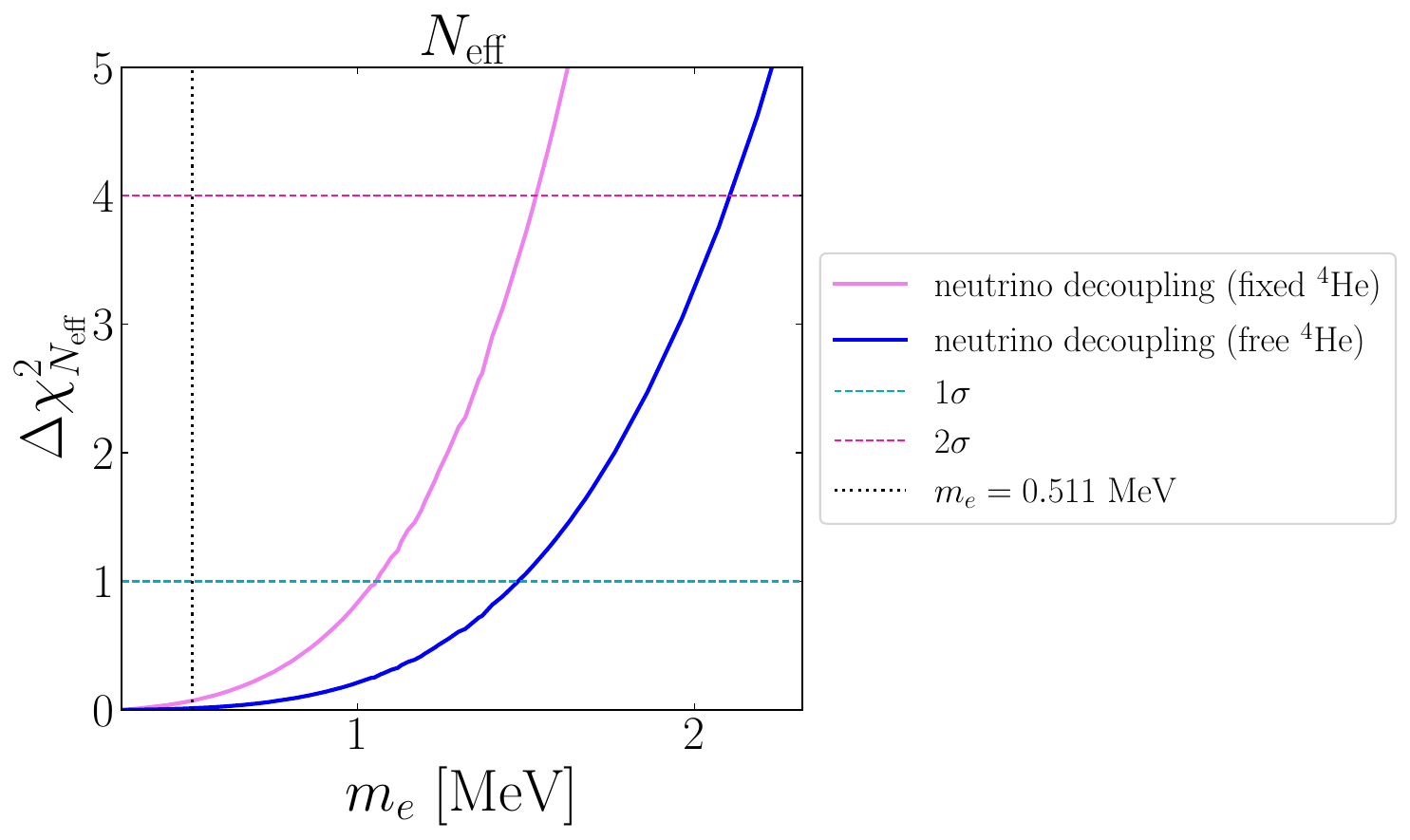}
\includegraphics[width=0.49\textwidth]{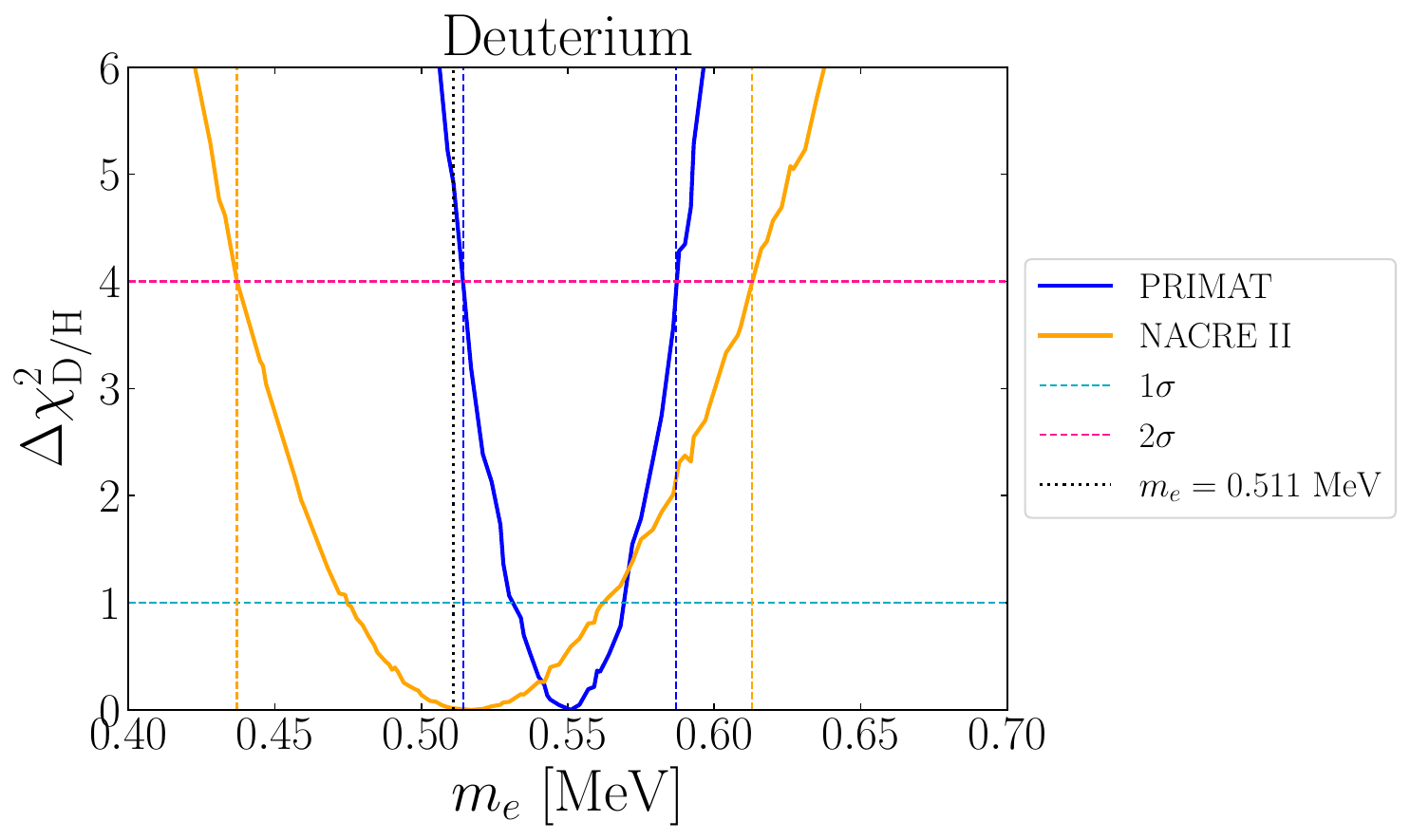}\\
\includegraphics[width=0.49\textwidth]{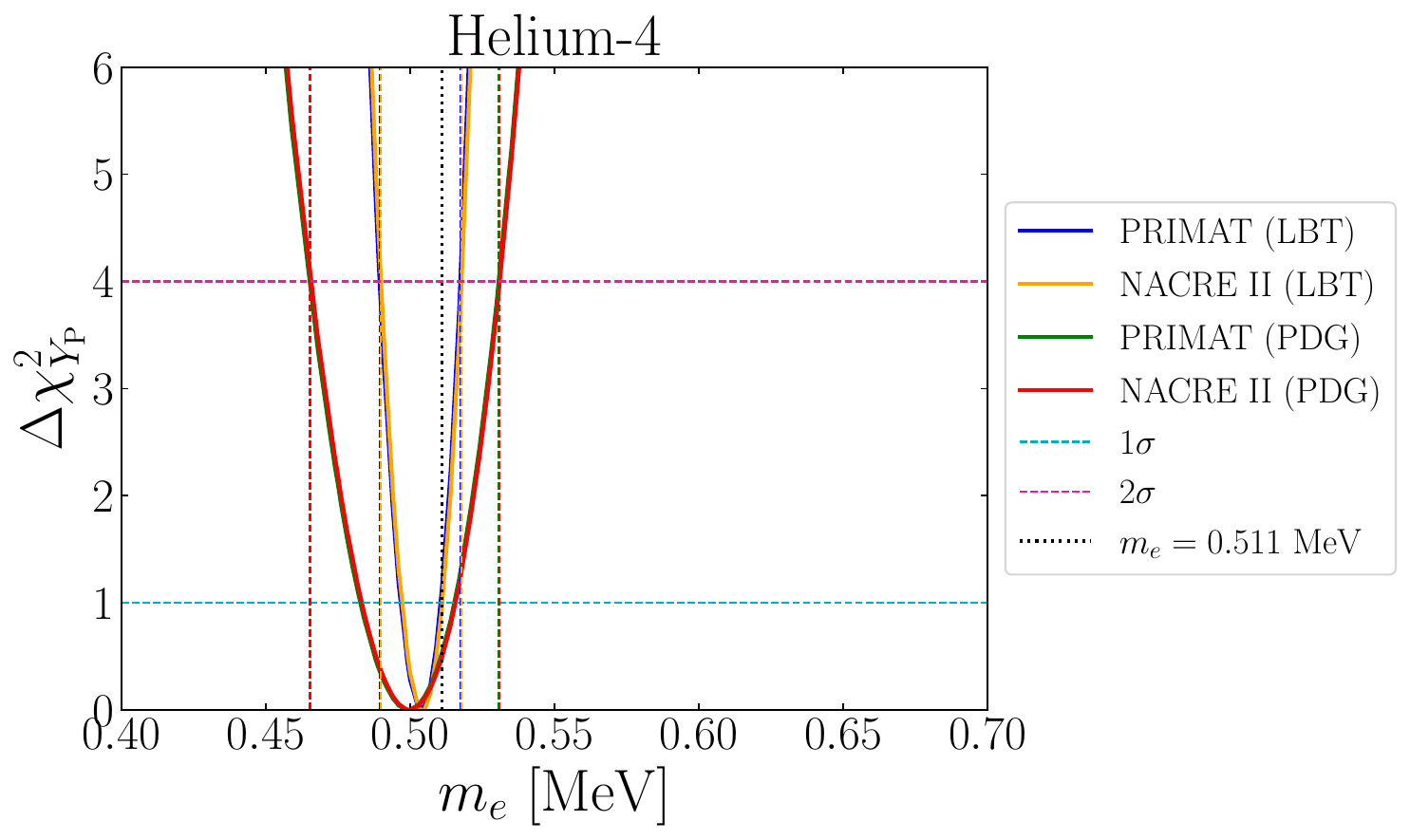}
\includegraphics[width=0.49\textwidth]{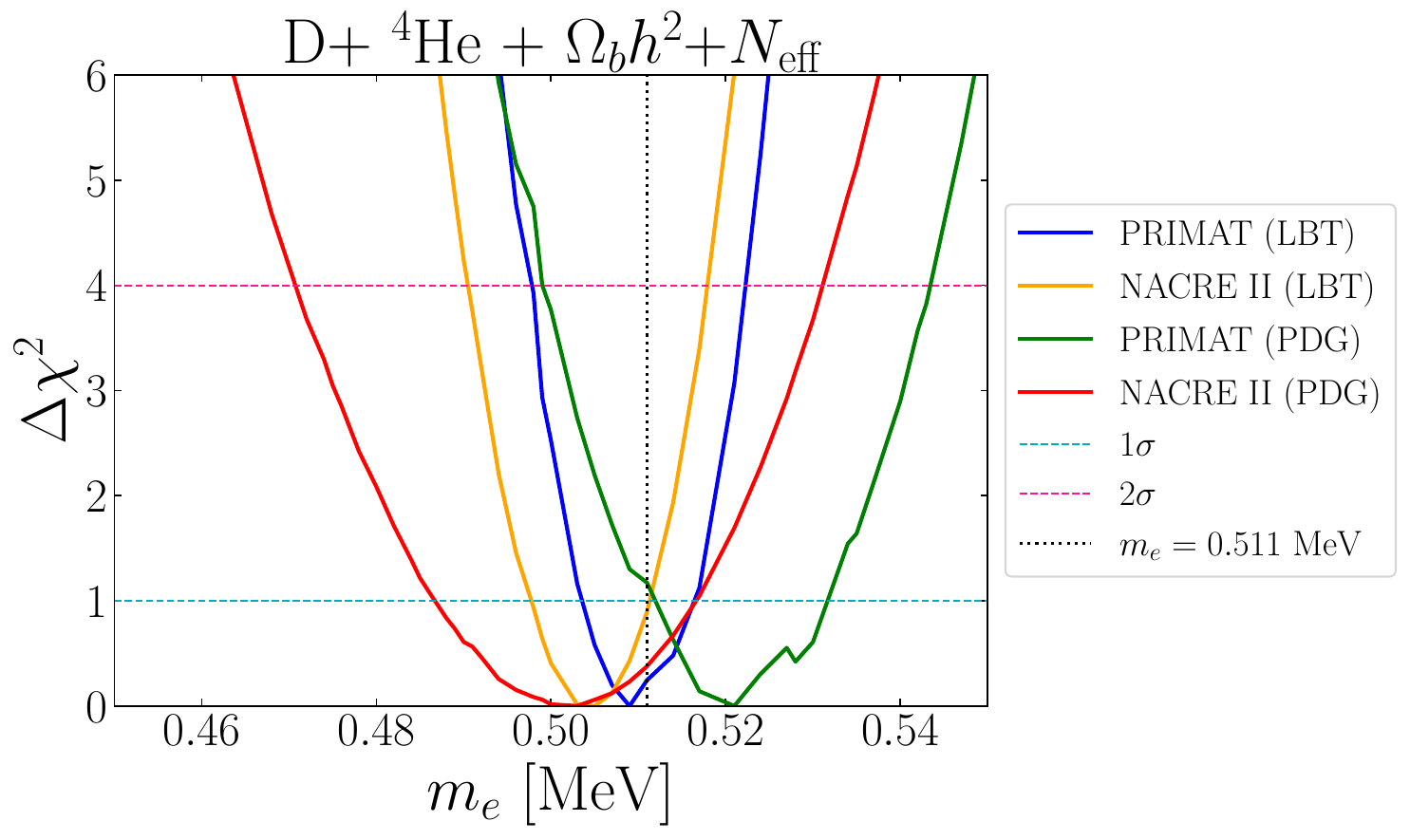}
\caption{$\Delta\chi^2$ analysis for \Neff, deuterium and helium-4, considered individually and combined. The total $\Delta\chi^2$ analysis includes the contribution of the $\Omega_b h^2$ parameter, too. The vertical dashed lines indicate the $1\sigma$ and $2\sigma$ confidence limits obtained for both the PRIMAT (blue and green) and NACRE II (orange and red) reaction-rate compilations.}
\label{fig:dchi}
\end{figure*}

\begin{figure*}[t]
    \centering
    \includegraphics[width=\textwidth]{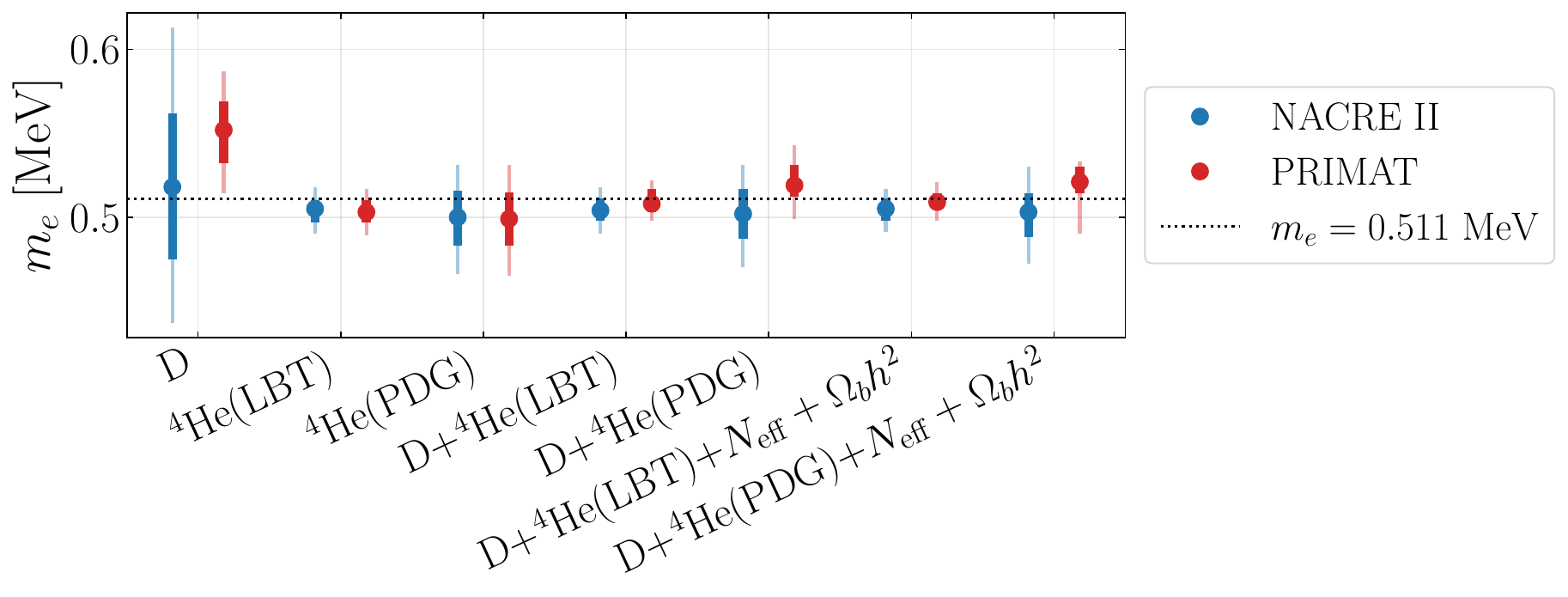}
    \caption{Constraints on $m_e$ derived from different observables. The figure shows the Table \ref{tab:values} results. Thick (thin) bars refer to $1\sigma$ ($2\sigma$) bounds.}
    \label{fig:me_constraints}
\end{figure*}

The individual constraints from $N_{\mathrm{eff}}$, deuterium, and helium-4 are shown in the top row panels and in the left bottom row panel of Fig.~\ref{fig:dchi}. The blue (orange) curves correspond to PRIMAT (NACRE II) and adopt the LBT \ce{^4He} abundance, while the green (red) curves correspond to PRIMAT (NACRE II) for the PDG \ce{^4He} value. The standard value $m_e=0.511~\MeV$ is shown by a vertical dotted black line. The corresponding confidence intervals are summarized in Table~\ref{tab:values} and illustrated in Fig.~\ref{fig:me_constraints}.

\paragraph{Constraint from $N_{\mathrm{eff}}$.}
The \Neff\ likelihood is identical for the PRIMAT and NACRE II analyses, since \Neff\ does not depend on nuclear reaction rates. Its dependence on $m_e$ is shown in Fig.~\ref{fig:NEFF_NUOVO}, and in the left panel in the top row of Fig.~\ref{fig:dchi} yields the corresponding bounds. We obtain one-sided upper limits
\begin{equation}
m_e < 1.055~\MeV \quad (1\sigma),\qquad
m_e < 1.530~\MeV \quad (2\sigma).\label{eq:constraint1}
\end{equation}
These constraints are derived assuming a fixed $Y_{\mathrm{P}}$. However, knowing that this value could change as $m_e$ varies, alternative constraints were derived using $N_{\mathrm{eff}}^{\rm obs}$ in equation~\eqref{eq:Neff}, in order to completely disengage the parameter's relation with BBN:
\begin{equation}
m_e < 1.477~\MeV \quad (1\sigma),\qquad
m_e < 2.104~\MeV \quad (2\sigma).\label{eq:constraint2}
\end{equation}
These bounds are less stringent than the ones derived from a fixed $Y_{\rm P}$ \Neff. To ensure consistency in the analysis, we considered the free $Y_{\rm P}$ \Neff~in the $\chi^2_{\mathrm{tot}}$ analysis discussed below. In both situations, The absence of a meaningful lower bound follows from the fact that decreasing $m_e$ delays $e^\pm$ annihilation to temperatures at which neutrinos are already decoupled; the annihilation then heats only the photon bath, and further reductions in $m_e$ do not appreciably change the entropy transfer. As a result, \Neff\ saturates close to its instantaneous-decoupling limit for sufficiently small $m_e$, and present data cannot discriminate among lower values. In contrast, increasing $m_e$ shifts $e^\pm$ annihilation to earlier times, while neutrinos are still coupled to the electromagnetic plasma; neutrinos then share the entropy release, increasing their energy density and thus \Neff. For sufficiently large $m_e$, the predicted \Neff\ becomes incompatible with the observational constraint, yielding the upper limits above.

\paragraph{Constraint from deuterium.}
The right panel in the top row of Figure~\ref{fig:dchi} shows that deuterium provides the strongest dataset dependence between NACRE II and PRIMAT. The NACRE II compilation yields the best agreement with the observed deuterium abundance but also the broadest allowed range, whereas PRIMAT produces a narrower interval with a best-fit value shifted to slightly higher $m_e$. This behavior has two main origins. First, the compilations differ in key reaction rates that process deuterium into \ce{^3H} and \ce{^3He}. NACRE II relies primarily on global fits to experimental data~\cite{Xu2013NACREII}, whereas PRIMAT imports and combines cross-section inputs from dedicated analyses~\cite{Pitrou_2018,Marcucci_2016,Leonard_2006} within a BBN-optimized framework. Second, NACRE II predicts a deuterium abundance closer to the observed value already at the standard electron mass $m_e=0.511~\MeV$, which improves the apparent consistency of the fit. The larger theoretical uncertainties typically associated with NACRE II broaden the confidence intervals, but they do not by themselves generate better agreement; rather, they enlarge the allowed region around the best-fit point.

\paragraph{Constraint from helium-4.}
The situation is qualitatively different for helium-4, as shown in the left panel of the bottom row of Fig.~\ref{fig:dchi}. Due to the higher precision of the $\ce{^4He}$ measurement, the confidence intervals are narrower when using the LBT \cite{Aver:2026dxv} abundance. The inferred bounds are nearly identical for PRIMAT and NACRE II  (Table~\ref{tab:values}), and this is expected: once the neutron abundance at the onset of nucleosynthesis is set by weak freeze-out and neutron decay, almost all surviving neutrons are incorporated into \ce{^4He}, making $Y_{\rm P}$ only weakly sensitive to details of the nuclear reaction network. In addition, the theoretical uncertainties on $Y_{\rm P}$ are comparable in the two compilations.


\paragraph{Combined constraint.}
We finally combine deuterium, helium-4, and \Neff\ in $\chi^2_{\rm tot}$; the resulting $\Delta\chi^2$ curves are shown in the right panel of the bottom row in Fig.~\ref{fig:dchi}, and the corresponding intervals are given in the last rows of Table~\ref{tab:values}. The combined constraints are driven primarily by helium, which dominates the error budget by being the most precise in constraining the electron mass $m_e$ during BBN. This is particularly evident in the results obtained using the latest measurement of the helium abundance from LBT \cite{Aver:2026dxv}, but it remains true even when considering the PDG helium value. The least constraining input is the free $Y_{\rm P}$ \Neff, whose dependence on $m_e$ is relatively mild over the relevant range compared to its current observational uncertainty  $\sigma_{N_{\mathrm{eff}},\rm exp}=0.29$.

Overall, the combined analysis yields constraints at the percent level on $m_e$ during the BBN epoch.
We notice that the helium observations alone point towards a $m_e$ value slightly below the expected 0.511~\MeV,
which remains at the limit of the $1\sigma$ constraints.
The combination with deuterium constraints obtained when considering the PRIMAT reaction network, shifts the $m_e$ central value closer to 0.511~\MeV.

\begin{table*}[t]
\centering
\resizebox{\textwidth}{!}{
\begin{tabular}{|c|c||c|c|c||c|c|c|}
    \hline
    Obs.&Nuc. Net.
    & \multicolumn{3}{c||}{$1\sigma$}
    & \multicolumn{3}{c|}{$2\sigma$}\\
    \hline
&
& $m_e$ range [MeV]
& $m_e$ [MeV] & $m_e/(0.511~\mathrm{MeV})$
& $m_e$ range [MeV]
& $m_e$ [MeV] & $m_e/(0.511~\mathrm{MeV})$\\
    \hline
    \hline
    $N_{\mathrm{eff}}$ (fixed \ce{^4He}) &
    &\multicolumn{2}{c|}{$ < 1.055$}
    & $<2.064$
    & \multicolumn{2}{c|}{$< 1.530$}
    & $<2.994$\\
    \hline
    $N_{\mathrm{eff}}$ (free \ce{^4He}) &
    &\multicolumn{2}{c|}{$ < 1.477$}
    & $<2.890$
    & \multicolumn{2}{c|}{$< 2.104$}
    & $<4.117$\\
    \hline
    \hline
\multirow{2}{*}{\ce{D}} & {NACRE II}
&$[0.475,0.562]$
&$0.518^{+0.044}_{-0.043}$& $1.014^{+0.086}_{-0.084}$
&$[0.437,0.613]$
& $0.518^{+0.095}_{-0.081}$&$1.014^{+0.185}_{-0.158}$\\
     & {PRIMAT}
     &$[0.532,0.569]$
     &$0.552^{+0.017}_{-0.020}$&$1.080^{+0.033}_{-0.039}$
     & $ [0.514,0.587]$
     & $0.552^{+0.035}_{-0.038}$&$1.080^{+0.068}_{-0.074}$ \\
    \hline
\multirow{2}{*}{\ce{^4He} (LBT)} & {NACRE II}
& $[0.497,0.511]$&$0.505^{+0.005}_{-0.008}$&$0.988^{+0.009}_{-0.015}$
& $[0.490,0.518]$
&$0.505^{+0.013}_{-0.015}$&$0.988^{+0.025}_{-0.029}$\\
                           & {PRIMAT}
                           & $[0.497,0.510]$
                           &$0.503^{+0.007}_{-0.006}$&$0.984^{+0.014}_{-0.012}$
                           &$[0.489,0.517]$
                           &$0.503 \pm 0.014$&$0.984 \pm 0.027$\\
    \hline
\multirow{2}{*}{\ce{^4He} (PDG)} & {NACRE II}
& $[0.483,0.516]$&$0.500^{+0.016}_{-0.017}$&$0.978^{+0.031}_{-0.033}$
& $[0.466,0.531]$
&$0.500^{+0.031}_{-0.034}$&$0.978^{+0.060}_{-0.066}$\\
                           & {PRIMAT}
                           & $[0.483,0.515]$
                           &$0.499\pm 0.016$&$ 0.976\pm0.031$
                           &$[0.465,0.531]$
                           &$0.499 ^{+0.032}_{-0.034}$&$0.976 ^{+0.062}_{-0.066}$\\
    \hline
\multirow{2}{*}{\ce{D}+\ce{^4He} (LBT)} & {NACRE II}
& $[0.498,0.511]$
& $0.504^{+0.007}_{-0.006}$&$0.986^{+0.013}_{-0.011}$
&$[0.490,0.518]$
& $0.504\pm{0.014}$&$0.986\pm{0.027}$\\
                      & {PRIMAT}
                      & $[0.503,0.517]$
                      & $0.508^{+0.009}_{-0.005}$ &$0.994^{+0.017}_{-0.009}$
                      & $[0.498,0.522]$
                      &$0.508^{+0.014}_{-0.010}$
                      &$0.994^{+0.027}_{-0.019}$\\
\hline
\multirow{2}{*}{\ce{D}+\ce{^4He} (PDG)} & {NACRE II}
& $[0.487,0.517]$
& $0.502\pm{0.015}$&$0.982\pm{0.029}$
&$[0.470,0.531]$
& $0.502^{+0.029}_{-0.032}$&$0.982^{+0.056}_{-0.062}$\\
                      & {PRIMAT}
                      & $[0.512,0.531]$
                      & $0.519^{+0.012}_{-0.007}$ &$1.015^{+0.023}_{-0.013}$
                      & $[0.499,0.543]$
                      &$0.519^{+0.024}_{-0.020}$
                      &$1.015^{+0.046}_{-0.039}$\\
\hline
\hline
\multirow{2}{*}{\ce{D}+\ce{^4He}(LBT)+$N_{\mathrm{eff}}$+$\Omega_bh^2$} & {NACRE II}
& $[0.498,0.511]$
& $0.505^{+0.006}_{-0.007}$&$0.988^{+0.011}_{-0.013}$
&$[0.491,0.517]$
& $0.505^{+0.012}_{-0.014}$&$0.988^{+0.023}_{-0.027}$\\
                      & {PRIMAT}
                      & $[0.505,0.514]$
                      & $0.509^{+0.005}_{-0.004}$ &$0.996^{+0.010}_{-0.008}$
                      & $[0.498,0.521]$
                      &$0.509^{+0.012}_{-0.011}$
                      &$0.996^{+0.023}_{-0.021}$\\
\hline
\multirow{2}{*}{\ce{D}+\ce{^4He}(PDG)+$N_{\mathrm{eff}}$+$\Omega_bh^2$} & {NACRE II}
& $[0.488,0.514]$
& $0.503^{+0.011}_{-0.015}$&$0.984^{+0.021}_{-0.029}$
&$[0.472,0.530]$
& $0.503^{+0.027}_{-0.031}$&$0.984^{+0.052}_{-0.060}$\\
                      & {PRIMAT}
                      & $[0.514,0.530]$
                      & $0.521^{+0.009}_{-0.007}$ &$1.019^{+0.017}_{-0.013}$
                      & $[0.500,0.543]$
                      &$0.521^{+0.022}_{-0.021}$
                      &$1.019^{+0.043}_{-0.041}$\\
\hline
\end{tabular}
}
\caption{$m_e$ constraints at $68.27\%$ C.L.\ ($1\sigma$) and $95.45\%$ C.L.\ ($2\sigma$). We report the corresponding best-fit values and their ratios to the laboratory value $m_e=0.511~\MeV$. All entries are in MeV. For each observable (except \Neff), the first row corresponds to NACRE II and the second to PRIMAT.}
\label{tab:values}
\end{table*}

\section{\label{sec:concl}Conclusions}
Early-Universe physics provides a powerful consistency test of the standard cosmological model and, more broadly, a unique laboratory for probing possible time variations of fundamental parameters. In this work we combined measurements of primordial light-element abundances from Big Bang Nucleosynthesis (BBN) with the effective number of relativistic species inferred from neutrino decoupling to constrain the value of the electron mass in the early Universe, allowing it to differ from the present laboratory value $m_{e,0}=0.511~\MeV$. This is a particularly informative epoch, spanning from a few seconds (neutrino decoupling) to a few minutes (BBN) after the Big Bang, for which no direct terrestrial information on $m_e$ is available.

The first part of our analysis solved the coupled differential equations \eqref{eq:dzodxp} and \eqref{eq:dz_neutrino}, which describe the evolution of the comoving photon and neutrino temperatures. We implemented these equations in a modified version of \texttt{NUDEC\_BSM}, generalized to a nonstandard electron mass and expressed in dimensionless variables. This procedure yields \Neff\ as a function of $m_e$ (Fig.~\ref{fig:NEFF_NUOVO}). For sufficiently large $m_e$, $e^+e^-$ annihilation occurs earlier, while neutrinos are still efficiently coupled to the electromagnetic plasma; neutrinos then share the released entropy, increasing their energy density and driving \Neff\ up to values as large as $\sim 11.55$ in the tightly coupled limiting case.

In the second part of the work, we studied the impact of a varying $m_e$ on BBN using the \texttt{PRyMordial} code, modified to propagate changes in $m_e$ through the weak sector and thermodynamics. A larger electron mass suppresses charged-current processes by reducing the available phase space, including neutron $\beta$ decay $n\to p+e^-+\bar{\nu}_e$, thereby lowering the neutron decay rate and increasing the effective neutron lifetime. These microphysical changes translate into a larger neutron abundance at the onset of nucleosynthesis, enhancing the production of neutron-rich nuclei, most notably \ce{^4He}. Correspondingly, the final abundance of free protons decreases slightly. For smaller $m_e$, nucleosynthesis begins earlier in time: although neutrons decay more rapidly, the higher temperature and density allow the reaction network to turn on earlier than in the standard scenario. 

We also isolated the role of the expansion history by implementing in \texttt{PRyMordial} only
$\Delta N_{\mathrm{eff}} \equiv N_{\mathrm{eff}}(m_e)-N_{\mathrm{eff}}(m_e=0.511~\MeV)$,
while keeping the weak rates fixed. Over the range explored here, the resulting shifts in the light-element abundances are negligible compared to the full effect, confirming that the dominant sensitivity to $m_e$ arises from direct modifications of the weak rates, neutron lifetime, and plasma thermodynamics rather than from the associated change in the expansion rate.

We performed the BBN analysis using both the PRIMAT and NACRE II nuclear reaction-rate compilations available in \texttt{PRyMordial}. The resulting differences are most evident for deuterium, due to its high sensitivity to the first nuclear reactions, as confirmed when considering this nucleus as a function of $m_e$ and $\Omega_b h^2$. In particular, PRIMAT predicts systematically lower abundances than NACRE II. This discrepancy propagates into the inferred preferred values of $m_e$ when deuterium is included in the fit and remains visible in the corresponding $\chi^2$ analyses, although at a reduced significance once helium-4 is included. Lithium-7 was excluded from the statistical analysis because the long-standing ``lithium problem'' would introduce additional, poorly controlled systematics.

Finally, we derived constraints on $m_e$ through a $\chi^2$ analysis of individual observables and of the combined statistic $\chi^2_{\rm tot}$. Using the minima of the $\Delta\chi^2$ curves as best-fit values, we obtain for the LBT \ce{^4He} value 
$m_e = 0.505^{+0.006}_{-0.007}~\MeV$ at $68.27\%$ C.L.\ and $m_e = 0.505^{+0.012}_{-0.014}~\MeV$ at $95.45\%$ C.L.\ with NACRE II, and
$m_e = 0.509^{+0.005}_{-0.004}~\MeV$ at $68.27\%$ C.L.\ and $m_e = 0.509^{+0.012}_{-0.011}~\MeV$ at $95.45\%$ C.L.\ with the PRIMAT network.
When using the PDG~\cite{ParticleDataGroup:2024cfk}  \ce{^4He} value, the obtained bounds are $m_e = 0.503^{+0.011}_{-0.015}~\MeV$ at $68.27\%$ C.L.\ and $m_e = 0.503^{+0.027}_{-0.031}~\MeV$ at $95.45\%$ C.L.\ with NACRE II, and
$m_e = 0.521^{+0.009}_{-0.007}~\MeV$ at $68.27\%$ C.L.\ and $m_e = 0.521^{+0.022}_{-0.021}~\MeV$ at $95.45\%$ C.L.\ with the PRIMAT network.
These bounds show that helium-4 observations dominate the final uncertainty. This is notably evident when using the \ce{^4He} LBT value in the analysis, because the \ce{^4He} better compensates for the deuterium reaction dependence.
In both cases, the shift in the preferred $m_e$ depends on the adopted reaction network: the NACRE II best fit lies slightly below $m_{e,0}$ (at the level of $\sim 1\sigma$), whereas the PRIMAT result is consistent with $m_{e,0}$.
Overall, both compilations are suitable for this analysis and provide complementary information, given their different nuclear inputs and uncertainty assignments.

Our combined constraints determine $m_e$ at the BBN epoch at the percent level, with an uncertainty of order $\sim 1.2\%$ at most (considering the new LBT \ce{^4He} observation \cite{Aver:2026dxv} in the analysis). While a number of works have constrained $m_e$ at later epochs in cosmic history~\cite{Seto_2023,Hart2017,Landau_2008,Toda:2024ncp}, the results presented here establish comparably stringent bounds at significantly earlier times, during BBN. In particular, we find that the electron mass at MeV temperatures must match its present laboratory value to within $\mathcal{O}(1\%)$, placing strong limits on any cosmological evolution of $m_e$ over an exceptionally wide range of cosmic time.

\begin{acknowledgments}
    M.G., N.F\ and S.G.\ are supported by the Research grant TAsP (Theoretical Astroparticle Physics) funded by Istituto Nazionale di Fisica Nucleare (INFN).
    S.G.\ also acknowledges financial support through the Ram\'on y Cajal contract RYC2023-044611-I funded by MICIU/AEI/10.13039/501100011033 and FSE+, and by the Spanish grants PID2023-147306NB-I00 and CEX2023-001292-S (MCIU/AEI/10.13039/501100011033).
\end{acknowledgments}

\appendix
\section{\label{sec:appendixA}
Useful expressions}

In this appendix we provide the definitions needed to rewrite the continuity equation~\eqref{eq:continuity} in the form of Eq.~\eqref{eq:dzodxp}. We begin with the standard comoving (dimensionless) energy density and pressure for thermal species.

\paragraph{Photons.}
For photons,
\begin{align}
   \bar{\rho}_\gamma &\equiv a^4 \rho_\gamma = \frac{\pi^2}{15}\,z^4,\\
   \bar{P}_\gamma &\equiv a^4 P_\gamma = \frac{\pi^2}{45}\,z^4,
\end{align}
so that
\begin{align}
    x'\,\frac{d\bar{\rho}_\gamma}{dx'} &= \frac{4\pi^2}{15}\,x' z^3 \frac{dz}{dx'},\\
    \bar{\rho}_\gamma-3\bar{P}_\gamma &= 0.
\end{align}

\paragraph{Electrons and positrons.}
For the $e^\pm$ plasma,
\begin{align}
  \bar{\rho}_e &\equiv a^4 \rho_e
  = \frac{g_e}{2\pi^2}\int_0^\infty dy\;
  \frac{y^2\,\sqrt{y^2 + k^2 x'^2}}{e^{\sqrt{y^2 + k^2 x'^2}/z}+1},\\
  \bar{P}_e &\equiv a^4 P_e
  = \frac{g_e}{6\pi^2}\int_0^\infty dy\;
  \frac{y^4}{\sqrt{y^2 + k^2 x'^2}}\,
  \frac{1}{e^{\sqrt{y^2 + k^2 x'^2}/z}+1}.
\end{align}
It is convenient to introduce the change of variables
\begin{equation}
\begin{cases}
\tau \equiv \dfrac{kx'}{z},\\[4pt]
\omega \equiv \dfrac{y}{z},\\[4pt]
dy = z\,d\omega,
\end{cases}
\label{change}
\end{equation}
together with the auxiliary functions
\begin{align}
    J_2(\tau) &= \frac{1}{\pi^2} \int_0^\infty d\omega\;
    \omega^2\,\frac{e^{\sqrt{\omega^2 + \tau^2}}}{\left(1+e^{\sqrt{\omega^2 + \tau^2}}\right)^2},\\
    J_4(\tau) &= \frac{1}{\pi^2} \int_0^\infty d\omega\;
    \omega^4\,\frac{e^{\sqrt{\omega^2 + \tau^2}}}{\left(1+e^{\sqrt{\omega^2 + \tau^2}}\right)^2}.
\end{align}

\paragraph{Neutrinos.}
For neutrinos we write
\begin{align}
\bar{\rho}_{\nu}
&\equiv a^4\rho_\nu
= \frac{1}{\pi^2}\int_0^\infty dy\; y^3 \sum_{\alpha=e,\mu,\tau} f_\alpha(y)\notag\\
&= \sum_{\alpha}\frac{7}{8}\frac{\pi^2}{15}\,z_\alpha^4,
\end{align}
where $f_\alpha$ denotes the neutrino distribution function for flavor $\alpha$.
This implies
\begin{align}\label{nu}
x'\,\frac{d\bar{\rho}_\nu}{dx'}
&= \frac{x'}{\pi^2}\int_0^\infty dy\; y^3 \sum_\alpha \frac{d f_\alpha}{dx'}\notag\\
&= 4x'\sum_\alpha \frac{7}{8}\frac{\pi^2}{15}\,z_\alpha^3\,\frac{dz_\alpha}{dx'}
= 4x'\sum_\alpha \frac{\bar\rho_{\nu_\alpha}}{z_\alpha}\,\frac{dz_\alpha}{dx'},\\
\bar{\rho}_\nu-3\bar{P}_\nu &= 0.
\end{align}
In the last equality we assumed that $f_\alpha$ is a Fermi--Dirac distribution characterized by a temperature $T_{\nu_\alpha}$ (equivalently $z_\alpha\equiv aT_{\nu_\alpha}$), so that $\bar\rho_{\nu_\alpha}$ depends on $x'$ only through $z_\alpha$.
By inserting the photon, $e^\pm$, and neutrino contributions into Eq.~\eqref{eq:continuity} and performing straightforward algebra, one obtains the explicit expression for $dz/dx'$ given in Eq.~\eqref{eq:dzodxp}.

\paragraph{Finite-temperature QED corrections.}
The functions $G_1$ and $G_2$ appearing in Eq.~\eqref{eq:dzodxp} arise from finite-temperature QED corrections at second order~\cite{Bennett:2019ewm}:
\begin{align}
    G_1(x',z) &= \frac{e^2k}{2} \left[
    \frac{1}{\tau}\left(\frac{K_2}{3}+2K_2^2-\frac{J_2}{6}-J_2K_2\right) \right. \\
     & \left. + \frac{K_2'}{6}-K_2K_2' + \frac{J_2'}{6}+J_2'K_2+J_2K_2'
    \right], \nonumber \\
    G_2(x',z) &= 2e^2\left(\frac{K_2^2}{2}-\frac{K_2}{6}-\frac{J_2}{6}-K_2J_2\right) \\
    &- \frac{e^2\tau}{2}\left(
    K_2K_2' - \frac{K_2'}{6} - \frac{J_2'}{6} - K_2'J_2 - K_2J_2'
    \right). \nonumber
\end{align}
Here we used
\begin{equation}
    K_2(\tau) = \frac{1}{\pi^2}\int_{0}^{\infty}d\omega\;
    \frac{\omega^2}{\sqrt{\omega^2+\tau^2}}\,
    \frac{1}{e^{\sqrt{\omega^2+\tau^2}}+1},
\end{equation}
where $\omega$ and $\tau$ are defined in Eq.~\eqref{change}. A prime denotes a derivative with respect to $\tau$. For compactness we do not explicitly indicate the $\tau$ dependence of $J_2$, $K_2$, and their derivatives elsewhere in the text.

\bibliographystyle{apsrev4-1}   
\bibliography{main}

\end{document}